\documentclass[11pt]{article}

\usepackage{graphicx}
\usepackage[margin=1.5cm]{geometry}
\usepackage{subcaption}
\usepackage{afterpage}

\newcommand{\avg}[1]{\left\langle #1\right\rangle}
\newcommand{\cov}[2]{\mathrm{Cov}\left(#1, #2\right)}
\newcommand{\var}[1]{\mathrm{Var}\,#1}

\pdfoutput=1

\usepackage{jcap2} 

\title{Hubble Diagrams in Statistically Homogeneous, Anisotropic Universes}

\author{Theodore Anton and}
\author{Timothy Clifton}

\affiliation{Department of Physics \& Astronomy, Queen Mary University of London, UK}

\emailAdd{t.j.anton@qmul.ac.uk; t.clifton@qmul.ac.uk}

\abstract{We consider the form of Hubble diagrams that would be constructed by observers in universes that are homogeneous but anisotropic, when averaged over suitably large length-scales. This is achieved by ray-tracing in different directions on the sky in families of exact inhomogeneous cosmological solutions of Einstein's equations, in order to determine the redshifts and luminosity distances that observers in these space-times would infer for distant astrophysical objects. We compare the results of this procedure to the Hubble diagrams that would be obtained by direct use of the large-scale-averaged anisotropic cosmological models, and find that observables calculated in the averaged model closely agree with those obtained from ray-tracing in all cases where a statistical homogeneity scale exists. In contrast, we find that in cosmologies with spaces that contain no statistical homogeneity scale that Hubble diagrams inferred from the averaged cosmological model can differ considerably from those that observers in the space-time would actually construct. We hope that these results will be of use for understanding and interpreting recent observations that suggest that large-scale anisotropy may have developed in the late Universe.}

\begin{document}

\maketitle

\flushbottom


\section{Introduction}

Hubble diagrams describe the relationship between the redshift of light received from distant sources, and the luminosity distance to them. They are of fundamental importance in cosmology, as they have played a crucial historical role, and underpin a wide array of cosmological observables. Of course, Hubble diagrams are typically interpreted within the homogeneous and isotropic Friedmann-Lema\^{i}tre-Robertson-Walker (FLRW) models of the Universe, and within this class of models can be used to determine the isotropic Hubble rate, $H_0$, and deceleration parameter $q_0$. However, they can also be constructed in anisotropic cosmological models. In such cases, the luminosity distance as a function of redshift depends on direction in space, and one could imagine constructing Hubble diagrams along certain lines of sight, so that $H_0$ and $q_0$ become functions on the sky. This issue has been brought to the fore by the apparent direction-dependence of dipole asymmetries in the CMB and matter distribution, which has led some to ask whether the late Universe might in fact be best described as being anisotropic on large scales \cite{Aluri_2023}. 

In a previous paper we considered how such an anisotropy might potentially emerge from structure formation \cite{Anton_2023}. In this work, we consider what the observational consequences of such a scenario might be, focusing in particular on how Hubble diagrams might be constructed in cosmologies that have emergent large-scale anisotropy in their expansion, and how the values of $H_0$ and $q_0$ inferred from such diagrams might be related to the average expansion and shear of space itself. In order to investigate this possibility we use the Sachs formalism for propagating bundles of rays of light in general space-times, as well as our general framework for incorporating the direction-dependent back-reaction of inhomogeneities on the large-scale properties of space.
Our approach is entirely relativistic and non-perturbative, and will therefore allow us to test explicitly whether the optical properties of inhomogeneous cosmological models can be described by an anisotropic average, even in situations where the inhomogeneities are of very large scale and/or amplitude.

The problem of calculating luminosity distances in inhomogeneous cosmologies is far from a new one \cite{Dyer_1973}, with numerous studies having been performed in (for example) Swiss cheese models \cite{Koksbang_2020_a,Koksbang_2020_b,Brouzakis:2006dj,Brouzakis:2007zi,Biswas:2007gi,Vanderveld:2008vi}, Lema\^{i}tre-Tolman-Bondi and Szekeres cosmologies \cite{Marra:2007pm,Marra:2007gc,Bolejko_2008,Clifton:2009nv,Bolejko:2008xh,Bolejko:2010eb,Peel:2014qaa,Koksbang:2017arw}, Lindquist-Wheeler models \cite{LW1, LW2, Clifton:2011mt,Liu:2015bya}, post-Newtonian cosmologies \cite{Sanghai:2017yyn}, N-body simulations \cite{Koksbang:2015ima,Koksbang:2023wez, adamek}, models constructed using numerical relativity \cite{Heinesen_2022,macpherson2021luminosity,Macpherson:2022eve}, and cosmographic analyses that construct the low-$z$ Hubble diagram in a generalised way \cite{Heinesen_2021, kalbouneh2024cosmography, maartens2024covariant,heincos}.
In this work, we will extend this field by investigating the extent to which Hubble diagrams in inhomogeneous universes can be accurately described by a large-scale averaged model that is anisotropic. We will study not only the all-sky average of the Hubble diagram (i.e. the monopole), but also the full variation of the luminosity distance function across the skies of many observers. This is made possible by the framework we built in Ref. \cite{Anton_2023}, which is an extended version of the spatial averaging procedure of Buchert \cite{Buchert_2000}. We find that our formalism can account well for that full variation, as long as the average model is allowed to be anisotropic, and as long as the spatial averaging is done on an appropriate foliation of the space-time.

The family of space-times we choose to consider for our study are dust-filled and plane symmetric. 
These solutions are discussed in detail in Section \ref{sec:plane_symm}, after we discuss our averaging formalism in Section \ref{sec:averaging}, and the formalism we use to calculate distance measures in Section \ref{sec:null_formalism}. In Section \ref{sec:sinusoidal} we bring these techniques together in the context of inhomogeneous and anisotropic space-times with zero back-reaction. We show that in that case there is a unique choice of homogeneous model corresponding to the scalar averages, and that within this class of models the averaged geometry permits a very good understanding of the Hubble diagrams of observers within it. We follow the same procedure in Sections \ref{sec:farnsworth} and \ref{sec:linear}, for space-times with non-zero back-reaction. Section \ref{sec:farnsworth} deals with a universe that is homogeneous, but where the surfaces of homogeneity are tilted with respect to the matter distribution. In contrast, Section \ref{sec:linear} considers an inhomogeneous and anisotropic space-time where backreaction is small but non-zero. We conclude in Section \ref{sec:discussion}. Throughout this paper, we use units such that $c = 8\pi G = 1$, and adopt the metric signature $\left(-,+,+,+\right)$. Latin indices from the beginning of the alphabet $a, b, c, ..$ denote space-time indices, whereas those from the middle of the alphabet $i, j, k, ...$ are reserved for space only.

\section{Cosmological Averaging with Anisotropy}\label{sec:averaging}

In order to extract the large-scale cosmological behaviour of an inhomogeneous space-time one requires an averaging, or coarse-graining, procedure. For this we use the spatial averaging procedure pioneered by Buchert \cite{Buchert_2000}, originally used in the context of the $1$+$3$-decomposition of Ehlers and Ellis \cite{Ellis_1999}. This is one of the most well-established averaging formalisms available, and can be applied to any covariantly defined scalars $S$ in a three-dimensional spatial domain $\mathcal{D}$ of constant time $t$ as follows: 
\begin{equation}\label{eq_Buchert_average}
    \avg{S}(t) := \frac{\int_{\mathcal{D}} \mathrm{d}^3 x \, \sqrt{^{(3)}g(t,x^i)} \, S(t,x^i)}{V_{\mathcal{D}}}     \, ,
\end{equation}
where $^{(3)}g_{ab}=g_{ab} +n_an_b $ is the induced spatial metric in $\mathcal{D}$, where the spatial volume of the domain is given by $V_{\mathcal{D}}=\int_{\mathcal{D}} \mathrm{d}^3 x \, \sqrt{^{(3)}g(t,x^i)}$, and where $n_a = - N(t,x^i)\, \nabla_a t$ is a time-like normal to $\mathcal{D}$, with $N$ being the lapse function. The spatial volume, defined in this way, can be used to define an effective scale factor for $\mathcal{D}$ of the form
\begin{eqnarray}
    a_{\mathcal{D}}(t) := \left(\frac{V_{\mathcal{D}}(t)}{V_{\mathcal{D}}(t_0)}\right)^{1/3}\,.
\end{eqnarray}
Within this approach, the time-evolution of scalar averages can be computed using the commutation rule
\begin{equation}
    \partial_t \avg{S} - \avg{\partial_t S} = \avg{N\Theta S} - \avg{N\Theta} \avg{S} := \cov{N\Theta}{S}\,,
\end{equation}
where $\cov{\cdot}{\cdot}$ indicates a covariance of its two arguments. In the above equation $\Theta = \nabla_a n^a$ is the isotropic expansion associated with the congruence $n_a$, such that its (lapse-weighted) average satisfies $\avg{N \Theta} = 3\,\partial_t a_{\mathcal{D}}/a_{\mathcal{D}}$\,.

In Buchert's original approach \cite{Buchert_2000}, spatial averaging is carried out on the scalars that are required to describe a homogeneous and isotropic FLRW universe, resulting in the following averaged versions of the Raychaudhuri and Hamiltonian constraint equations:
\begin{eqnarray}\label{eq_Buchert_1}
    3\left(\frac{\partial_t a_{\mathcal{D}}}{a_{\mathcal{D}}}\right)^2 &=& \avg{\rho} - \frac{1}{2}\avg{^{(3)}R} - \frac{1}{2}\mathcal{B}\,,\\
    \label{eq_Buchert_2} 3 \frac{\partial_{t}^2 a_{\mathcal{D}}}{a_{\mathcal{D}}} &=& -\frac{1}{2}\avg{\rho} + \mathcal{B}\,,
\end{eqnarray}
where we have taken the matter content to be pressureless dust, and set the lapse equal to unity and the cosmological constant to zero. Here $\rho = T_{ab} n^a n^b$ is the local energy density of matter, and $^{(3)}R$ is the Ricci scalar curvature of the three-spaces orthogonal to $n^a$\,.
The term $\mathcal{B}$, often referred to as the `kinematical back-reaction scalar', is given by
\begin{equation}
    \mathcal{B} = \frac{2}{3}\left(\avg{\Theta^2} - \avg{\Theta}^2\right) - 2\avg{\sigma^2}\,,
\end{equation}
where $\sigma_{ab}$ is the shear tensor associated with $n_a$\,, and $\sigma^2 = \sigma_{ab}\sigma^{ab}/2$\,. Buchert's equations (\ref{eq_Buchert_1}-\ref{eq_Buchert_2}) provide a useful formalism for modelling the large-scale expansion of an inhomogeneous cosmological space-time, but do not allow one to calculate the anisotropy in the emergent expansion, which is the subject of the next section.

\subsection{Averaging to an anisotropic universe}\label{subsec:averaging_anisotropic}

If one has reason to believe that there may be substantial anisotropy, or simply wishes to test how well a certain anisotropic model fits observational data, then it is necessary to consider not only the average of the isotropic expansion, but also of a number of other quantities (as in Ref. \cite{Anton_2023}). The first step in this approach is to decompose the space-time not only with respect to the time-like vector $n^a$, but also a space-like vector field $m^a$, which could correspond (for example) to the axis of a cosmic dipole.  The general name for such a decomposition, and the equations that result, is the $1$+$1$+$2$-formalism \cite{Clarkson_2007}. 
It leads to a large set of covariantly-defined scalars, which may then be averaged in order to obtain an emergent anisotropic cosmology that is local rotationally symmetric (LRS) \cite{Ellis_1967}. A locally rotationally symmetric space-time is one in which every point has associated with it a single preferred spacelike direction, about which rotations leave the geometry unchanged. They do not need to be homogeneous in general, but when they are we will refer to them as `LRS cosmologies'.
Such LRS cosmologies exist within the Bianchi classification, and can be of types {\it I}, {\it II}, {\it III}, {\it V}, {\it VII} or {\it IX}, with types {\it I}, {\it V}, {\it VII} and {\it IX} containing FLRW as special cases.

Assuming that the matter content of the universe is well-described by pressureless dust, 
all the LRS space-times we are interested in can be described purely in terms of the expansion scalar, $\Theta$, the energy density $\rho\,$, and the following three additional covariantly-defined $1$+$1$+$2$-scalars: 
\begin{eqnarray}\label{eq_LRS_scalars}
    \phi &:=& D_a m^a\\
    \Sigma &:=& \sigma_{ab} m^a m^b = D_{\langle a} n_{b \rangle}\,m^a m^b \\
    \mathcal{E} &:=& E_{ab}m^a m^b = C_{acbd} n^c n^d m^a m^b \,,
\end{eqnarray} 
which correspond to the space-like expansion of $m^a$, the shear of $n^a$ and the electric part of the Weyl tensor, respectively. In the above, we have defined $D_a$ to be the covariant derivative projected into space-like hypersurfaces of constant $t$, and the notation $t_{\langle ab \rangle}$ denotes a rank-2 tensor $t_{ab}$ that has been made symmetric and trace-free as well as being projected into the hypersurfaces orthogonal to $n^a$.

After averaging, the LRS scalars $\left\lbrace \Theta, \rho, \phi, \Sigma, \mathcal{E}\right\rbrace$ should therefore be thought of as non-local, large-scale quantities that evolve according to
\begin{eqnarray}
&&\partial_t \avg{\Theta} + \frac{1}{3}\avg{\Theta}^2 + \frac{3}{2}\avg{\Sigma}^2 + \frac{1}{2}\avg{\rho} = \mathcal{B}_1 \label{first} \\
&&\partial_t \avg{\Sigma} + \frac{2}{3}\avg{\Theta}\avg{\Sigma} + \frac{1}{2}\avg{\Sigma}^2 + \avg{\mathcal{E}} = \mathcal{B}_3\\
&&\partial_t \avg{\phi} + \frac{1}{2}\left(\frac{2}{3}\avg{\Theta} - \avg{\Sigma}\right)\avg{\phi} = \mathcal{B}_6\\
&&\partial_t \avg{\mathcal{E}} + \left(\avg{\Theta} - \frac{3}{2}\avg{\Sigma}\right)\avg{\mathcal{E}} + \frac{1}{2}\avg{\rho}\avg{\Sigma} = \mathcal{B}_{12}\\
&&\partial_t\avg{\rho} + \avg{\Theta}\avg{\rho} = 0  \, ,\phantom{\frac{1}{2}} \label{eq_conservation_equation_avg}
\end{eqnarray} 
and which obey the constraints
\begin{eqnarray}
&&\frac{3}{2}\avg{\phi}\avg{\Sigma} = \mathcal{B}_2 \\
&&\frac{2}{9}\avg{\Theta}^2 - \frac{1}{2}\avg{\phi}^2 + \frac{1}{3}\avg{\Theta}\avg{\Sigma} -\avg{\Sigma}^2 - \frac{2}{3}\avg{\rho} - \avg{\mathcal{E}} = \mathcal{B}_5 \\
&&\frac{3}{2}\avg{\mathcal{E}}\avg{\phi} = \mathcal{B}_{14} \, . \label{last}
\end{eqnarray} 
These are a reduced set of equations, tailored to the case of plane-symmetric dust-filled space-times, and with the lapse function chosen to be $N=1$\,. For the full set of $15$ equations\footnote{The numbering of our $\mathcal{B}_i$ corresponds directly to the quantities in Ref. \cite{Anton_2023}, with the equations containing $\mathcal{B}_4$, $\mathcal{B}_{13}$, $\mathcal{B}_{15}$, and $\mathcal{B}_7$-$\mathcal{B}_{11}$ not required here.}, and a presentation and explanation of their origin, the reader is referred to Ref. \cite{Anton_2023}. 

The back-reaction scalars, $\mathcal{B}_i$, in Eqs. (\ref{first})-(\ref{last}) encode the contribution of inhomogeneities to the averaged equations, and are given by the following, in our current setting:
\begin{eqnarray*}
&&\mathcal{B}_1 = \frac{2}{3}\,\var{\Theta} - \frac{3}{2}\,\var{\Sigma} \\
&&\mathcal{B}_2 = \frac{2}{3}\avg{m^a D_a \Theta} - \avg{m^a D_a \Sigma} - \frac{3}{2}\,\cov{\phi}{\Sigma} \\
&&\mathcal{B}_3 = \frac{1}{3}\,\cov{\Theta}{\Sigma} - \frac{1}{2}\,\var{\Sigma} \\
&&\mathcal{B}_5 = \frac{1}{2}\,\var{\phi} - \frac{2}{9}\,\var{\Theta} + \var{\Sigma} - \frac{1}{3}\,\cov{\Theta}{\Sigma} \\
&&\mathcal{B}_6 = \frac{2}{3}\,\cov{\Theta}{\phi} + \frac{1}{2}\,\cov{\phi}{\Sigma} \\
&&\mathcal{B}_{12} = \frac{3}{2}\,\cov{\Sigma}{\mathcal{E}} - \frac{1}{2}\cov{\Sigma}{\rho} \\
&&\mathcal{B}_{14} = - \avg{m^a D_a \mathcal{E}} + \frac{1}{3}\,\avg{m^a D_a \rho} - \frac{3}{2}\cov{\phi}{\mathcal{E}} \, .
\end{eqnarray*}
It can be seen that these scalars are composed by averaging spatial gradients of scalars along the preferred direction $m^a$, as well from the variances (Var) and covariances (Cov) of various of the covariantly-defined scalars, where $\var S:=\cov{S}{S}$ for any scalar $S$.

\subsection{Homogeneous, anisotropic LRS cosmologies}\label{subsubsec:average_models}

Given a set of averaged scalars, one may wish to construct an homogeneous but anisotropic LRS cosmology out of them, in order to interpret the large-scale behaviour of the averaged space-time. In such a scenario, the existence of any non-zero back-reaction scalars would indicate that the inhomogeneities that have been averaged away are having an influence on the dynamics of the large-scale cosmology. For the simplest possible example, the averaged scalars $\avg{\Theta}$ and $\avg{\Sigma}$ can be used to specify a Bianchi type-{\it I} line-element of the form
\begin{equation}\label{eq_Bianchi_I}
    \mathrm{d}s^2_I = -\mathrm{d}t^2 + A^2(t)\, \mathrm{d}r^2 + B^2(t)\left(\mathrm{d}y^2 + \mathrm{d}z^2\right)\,,
\end{equation}
where $A(t)$ and $B(t)$ are scale factors that are related to our averaged scalars by
\begin{eqnarray}\label{eq_Bianchi_Theta_Sigma}
\avg{\Theta} =    \frac{\dot{A}}{A} + \frac{2\dot{B}}{B}  \qquad \mathrm{and} \qquad \avg{\Sigma}= \frac{2}{3}\left(\frac{\dot{A}}{A} - \frac{\dot{B}}{B}\right) \,.
\end{eqnarray}
Similarly, one can write down a Bianchi type-{\it V} line-element of the form
\begin{equation}\label{eq_Bianchi_V}
    \mathrm{d}s^2_V = -\mathrm{d}t^2 + A^2(t)\, \mathrm{d}r^2 + B^2(t) e^{-2\beta r}\left(\mathrm{d}y^2 + \mathrm{d}z^2\right) \, ,
\end{equation}
where $\avg{\Theta}$ and $\avg{\Sigma}$ are related to $A(t)$ and $B(t)$ as in the equations above, and where the spatial curvature parameter $\beta$ is given according to 
\begin{equation}\label{eq_Bianchi_V_beta}
    -\frac{6\beta}{A^2} - \frac{\dot{A}^2}{2A^2} - \frac{\dot{B}^2}{2B^2} + \frac{\dot{A}\dot{B}}{AB} = \big\langle {}^{(3)}R \big\rangle = 2\avg{\rho} - \frac{2}{3}\avg{\Theta}^2 + \frac{3}{2}\avg{\Sigma}^2\,,
\end{equation}
where $^{(3)}R$ is the Ricci scalar curvature of the hypersurfaces orthogonal to $n^a$. In the final equality here we have made use of the Hamiltonian constraint equation \cite{Ellis_1999} in order to relate $\beta$ to $\avg{\rho}$, $\avg{\Theta}$ and $\avg{\Sigma}$. In both cases we recover an FLRW model when $A(t)=B(t)$\,, as can be seen from the vanishing of the shear in this case.

\section{Redshifts and Distance Measures in General Space-Times}\label{sec:null_formalism}

To achieve our goal of testing how well emergent cosmological models describe observations in inhomogeneous and anisotropic universes we will require a general understanding of how to calculate redshifts and distance measures in General Relativity. There are three ingredients required for this: (i) a ray tracing procedure to determine the paths followed by photons, (ii) a formalism to describe null geodesic congruences, and (iii) a way to calculate distance measures from the properties of those congruences. These are explained below.

\subsection{Ray tracing}

In order to calculate redshifts and luminosity distances we need to know the trajectories of rays of light in a given space-time. Under the eikonal approximation these will be null geodesics with $k^b \nabla_b k_a = 0$, where $k^a$ is the tangent vector to the ray. Finding these paths can be achieved straightforwardly by constructing the Hamiltonian $\mathcal{H} := g^{ab}k_a k_b/2$, subject to the constraint $\mathcal{H}=0$, and using Hamilton's equations:
\begin{eqnarray}\label{eq_geodesic_equation}
    \frac{\mathrm{d}x^a}{\mathrm{d}\lambda} &=& \frac{\partial \mathcal{H}}{\partial k_a} = g^{ab}k_b\,, \\
    \nonumber \frac{\mathrm{d}k_a}{\mathrm{d}\lambda} &=& -\frac{\partial \mathcal{H}}{\partial x^a} = -\frac{1}{2}g^{bc}_{\ \ ,a}k_b k_c \,.
\end{eqnarray}
These equations will be integrated backwards in time from the observer to the source, by choosing $k^a$ to be past-directed. For the initial conditions, one requires a space-time location $x^a_{\rm obs}$, and a propagation direction $e_a^{\rm obs}$ along which rays arrive at the observer.  In defining $e_a^{\rm obs}$, we have decomposed the tangent vector $k^a$ with respect to the time-like vector $n^a$, such that $k^a = -E\left(n^a - e^a\right)$, where $e^a n_a = 0$ and $e^a e_a = 1$. The photon energy is then $E = k^a n_a$, and $e^a$ gives the direction of the ray in the space-like hypersurfaces orthogonal to $n^a$\,.

The initial direction of propagation $e_a^{\rm obs}$ is chosen by specifying the angles $\left(\theta_c, \phi_c\right)$ on the observer's celestial sphere. These angles pick out a space-like unit vector
\begin{equation}
\epsilon^i = -\left(\cos{\theta_c}, \sin{\theta_c}\cos{\phi_c}, \sin{\theta_c}\sin{\phi_c}\right)\, .
\end{equation}
Aligning $e_a^{\rm obs}$ with this unit vector, and using the null condition $k^a k_a=0$, is then sufficient to determine $k^a$ at the observer (up to the specific value of $E$). By varying the observing angles $\left(\theta_c, \phi_c\right)$, and the observer's location in space-time $x^a_{\rm obs}$, we can then calculate the path of any null geodesic. We can of course also calculate the redshift along any particular geodesic, using the following definition:
\begin{equation}\label{eq_redshift_def}
    1 + z := \frac{E(\lambda)}{E(0)} = \frac{k^a n_a\vert_{\lambda}}{k^a n_a \vert_{0}}\,.
\end{equation}
In order to construct the Hubble diagram, however, it is also necessary to consider congruences of such curves, which is what we will do in the next section.

\subsection{Geodesic null congruences}

By considering a family of many infinitesimally separated null geodesics with tangent $k^a$, one can construct a congruence that is orthogonal to a two-dimensional screen space, which defines the projection tensor $s_{ab} = g_{ab} + n_a n_b - e_a e_b$\,. By parallel-transporting the screen space basis along the null geodesic congruence, one can study how the properties of that congruence evolve, allowing us to study the optical properties of the space-time.

Expansion and shear of the screen space are defined by $\hat{\theta} := \frac{1}{2}\nabla_a k^a$ and $\hat{\sigma}_{ab} := \big(s_{(a}^{\ c} s_{b)}^{\ d} - \frac{1}{2} \,s_{ab}s^{cd}\big)\nabla_c k_d$, and the screen space itself is spanned by a pair of orthonormal space-like vectors $s_1^{\ a}$ and $s_2^{\ a}$, or equivalently by the complex null vector $s^a = \left(s_1^{\ a} - i s_2^{\ a}\right)/\sqrt{2}$ and its complex conjugate $\bar{s}^a$. 
The screen space projection tensor is related to these basis vectors by $s_{ab} = s^1_{\ a}s^1_{\ b} + s^2_{\ a}s^2_{\ b} = s_a\bar{s}_b + \bar{s}_a s_b\,$.
Since the null shear is symmetric, trace-free and fully projected into the (2-dimensional) screen space, it contains only two independent degrees of freedom. Therefore, it can be characterised entirely by a single complex scalar $\hat{\sigma} := - s^a s^b \hat{\sigma}_{ab}$\,. The evolution of $\hat{\theta}$ and $\hat{\sigma}$ along the null geodesics is governed by the Sachs equations \cite{schneider1992gravitational,Fleury:2015hgz,Bull_2012},
\begin{eqnarray}\label{sachs_old_1}
    \frac{\mathrm{d}\hat{\theta}}{\mathrm{d}\lambda} + \hat{\theta}^2 + \bar{\hat{\sigma}}\hat{\sigma} &=& -\frac{1}{2}R_{ab}k^a k^b \ = \: \Phi_{00}\, , \\
    \label{sachs_old_2} \frac{\mathrm{d}\hat{\sigma}}{\mathrm{d}\lambda} + 2\hat{\theta}\hat{\sigma} &=& C_{abcd}s^a k^b s^c k^d \ = \: \Psi_0 \,,
\end{eqnarray}
where $\Phi_{00} = -R_{ab} k^a k^b/2$ and $\Psi_0 = C_{abcd} s^a k^b s^c k^d$ are the Newman-Penrose scalars corresponding to Ricci focusing and Weyl lensing, respectively, with the null energy condition implying $\Phi_{00} \leq 0$\,.
These equations determine the evolution of the null expansion and shear, and are complemented by the following equation for the evolution of redshift
\begin{equation}\label{eq_dz_dlambda}
\frac{\mathrm{d}z}{\mathrm{d}\lambda} = \left(1+z\right)^2 H^{\parallel}(z)\,,
\end{equation}
where $H^{\parallel} = k^a k^b \nabla_a n_b /E^2 = \Theta/3 + \sigma_{ab} e^a e^b$ is the rate of expansion of space in the direction of the ray of light.

\subsection{Distance measures}

The measure of distance in which we are ultimately interested, for the construction of Hubble diagrams, is the luminosity distance, $d_L$. However, for a single observer seeing multiple astrophysical sources it is most straightforward to calculate the angular diameter distance $d_A$, which is related to the null expansion scalar $\hat{\theta}$ by
\begin{equation}
     \hat{\theta} = \frac{\mathrm{d}}{\mathrm{d}\lambda}\,\ln{d_A}\,.
\end{equation}
Working with $d_A$, the Sachs equations can be re-written as
\begin{eqnarray}\label{sachs1}
    \frac{\mathrm{d}^2}{\mathrm{d} \lambda^2}\, d_A &=& \left(\Phi_{00} - \bar{\hat{\sigma}}\hat{\sigma}\right) d_A \, \\
    \frac{\mathrm{d}}{\mathrm{d}\lambda}\, \left(\hat{\sigma}\, d_A^2\right) &=& \Psi_0\, d_A^2 \, , \label{sachs2}
\end{eqnarray}
with the initial conditions
\begin{eqnarray}
    d_A(0) = 0\,, \qquad \frac{\mathrm{d}}{\mathrm{d}z}d_A\Big\vert_0 = \frac{1}{H_0^{\parallel}}\qquad {\rm and} \qquad \hat{\sigma}(0) = 0\,,
\end{eqnarray}
where $H_0^{\parallel}$ denotes $H^{\parallel}$ evaluated at the observer's location. One may then use $d_A$ to determine the luminosity distance using Etherington's reciprocity theorem \cite{Etherington_1933,ellis2009republication,ellis2012relativistic}:
\begin{equation}\label{eq_reciprocity_theorem}
d_L = \left(1+z\right)^2 d_A\,,
\end{equation}
which allows a Hubble diagram to be constructed from the function $d_L(z)$ along any given line of sight. However, in practice it is often useful to display the Hubble diagram in terms of the distance modulus, defined by
\begin{equation}\label{eq_distance_modulus}
\mu := 5\log{\left(\frac{d_L}{\rm Mpc}\right)} + 25\,.
\end{equation}
We will use the distance modulus in much of the analysis we perform in Sections \ref{sec:sinusoidal}-\ref{sec:linear}.

\section{Plane-Symmetric Cosmological Models}\label{sec:plane_symm}

The next ingredient that we need in order to implement our formalism is a set of inhomogeneous cosmological models. Our intent is to average these models using the equations from Section \ref{sec:averaging}, and to calculate observables in both the averaged and un-averaged space-times using the approach outlined in Section \ref{sec:null_formalism}. An appropriate choice is provided by the family of plane-symmetric dust-filled cosmologies, which exhibit a single preferred space-like direction orthogonal to the homogeneous planes of symmetry, and which can exhibit arbitrary amounts of inhomogeneity along this direction. 
Such plane-symmetric space-times are necessarily LRS, as every point in each plane of symmetry has associated with it a 1-parameter isotropy group consisting of rotations within that plane \cite{Stephani_2003}.
 
The metric for plane-symmetric space-times can be written in the general form
\begin{equation}\label{eq_plane_symmetric_general_metric}
\mathrm{d}s^2 = -e^{2\nu(t)}\mathrm{d}t^2 + e^{2\lambda(t,r)}\mathrm{d}r^2 + R^2(t,r)\left(\mathrm{d}y^2 + \mathrm{d}z^2\right).
\end{equation}
This very general class of metrics includes the spatially-flat and negatively-curved FLRW, degenerate Kasner (with identical scale factors in the $y$ and $z$ directions), and vacuum Taub solutions as special cases. In this case, the solutions to Einstein's equations can be split into two distinct classes: (i) those with $\partial R/\partial r = 0\,$, and (ii) those with $\partial R/\partial r \neq 0$. Both of these classes allow for significant inhomogeneity, but only the second will turn out to have non-zero back-reaction scalars, $\mathcal{B}_i$\,. 
The $y$ and $z$ coordinates in Eq. (\ref{eq_plane_symmetric_general_metric}) label points in the planes of symmetry, while $t$ and $r$ correspond to time and space directions orthogonal to those planes. 
For these models, a natural choice for unit vectors in the preferred time-like and space-like directions is prescribed by $n_a = -\delta_a^{\ t}$ and $m_a = \exp \{ \lambda(t,r) \}\,\delta_a^{\ r}\,$, and the scalars $\left \lbrace \Theta,\rho,\phi,\Sigma,\mathcal{E}\right\rbrace\,$ are naturally all functions of $t$ and $r$ only. Plane-symmetric space-times are therefore very well-suited for study within our formalism, and for the remainder of this section we will outline the form the metric functions $\nu(t)$, $\lambda(t,r)$ and $R(t,r)$ must take in order to be solutions to Einstein's equations.

\subsection{Dust-filled solutions with $R' = 0$}

When $R' = 0$, where a prime denotes partial differentiation with respect to $r$, one can re-define the time coordinate such that $R(t) = t$\,. The metric functions can then be expressed as \cite{Stephani_2003, Clifton_2019}
\begin{equation}
    e^{2\nu(t)} = \frac{t}{t_0} \quad {\rm and} \quad e^{2\lambda(t,r)} = \frac{t_0}{t}\left[c_1(r) \left(\frac{t}{t_0}\right)^{3/2} + c_2(r) \right]\,,
\end{equation}
where $c_1(r)$ and $c_2(r)$ are arbitrary functions of $r$, and $t_0$ is a constant with units of time.
By a further re-definition of the time coordinate, $t \longrightarrow t_0 \left({3t}/{2t_0}\right)^{2/3}$, we can set $\nu=0$. Finally, the factors of $3/2$ and $t_0$ can be absorbed into $c_1(r)$ and $c_2(r)$, to end up with the general solution in the form
\begin{equation}\label{Rprime=0_metric}
\mathrm{d}s^2 = -\mathrm{d}t^2 + \left[c_1(r) \left(\frac{t}{t_0}\right)^{2/3} + c_2(r)\left(\frac{t}{t_0}\right)^{-1/3}\right]^2 \mathrm{d}r^2 + \left(\frac{t}{t_0}\right)^{4/3}\left(\mathrm{d}y^2 + \mathrm{d}z^2\right)\,.
\end{equation} 
The line-element above corresponds to the Einstein-de Sitter solution when $c_2$ vanishes, and the degenerate Kasner solution when $c_1$ vanishes. The $1$+$1$+$2$-scalars in this case are given by
\begin{eqnarray}\label{eq_R'=0_scalars}
\Theta = \frac{2tc_1+ t_0 c_2}{t\left(tc_1+ t_0 c_2\right)}\,, \:
\rho = \frac{4c_1}{3t\left(tc_1+ t_0 c_2\right)}\,, \:
\Sigma =  -\frac{2t_0 c_2}{3t\left(tc_1+ t_0 c_2\right)}\,, \:
\mathcal{E} =  -\frac{4t_0 c_2}{9t^2\left(tc_1+ t_0 c_2\right)} \, , \:
\end{eqnarray}
and $\phi = 0$.
Due to the plane symmetry, averages over space-like domains reduce to ratios of one-dimensional integrals over the specified range of the $r$ coordinate, such that
\begin{equation}\label{eq_R'=0_average_definition}
\avg{S}(t) \ := \ \frac{\int_{\mathcal{D}} \mathrm{d}^3 x \, \sqrt{^{(3)}g(t,r)} \, S(t,r)}{\int_{\mathcal{D}} \mathrm{d}^3 x \, \sqrt{^{(3)}g(t,r)}} \ = \ \frac{\int_{r_{\mathrm{min}}}^{r_{\mathrm{max}}} \mathrm{d}r \, \left(t c_1(r) + t_0 c_2(r)\right)\,S(t,r)}{\int_{r_{\mathrm{min}}}^{r_{\mathrm{max}}}\mathrm{d}r \, \left(t c_1(r) + t_0 c_2(r)\right)}\, ,
\end{equation}
for any scalar $S=S(t,r)$. Evaluating the back-reaction scalars in Section \ref{subsec:averaging_anisotropic}, one finds that for any choice of the functions $c_1(r)$, $c_2(r)$, and for any interval $\left(r_{\mathrm{min}},r_{\mathrm{max}}\right)$, all $\mathcal{B}_i=0$, as long as $c_1(r)$ and $c_2(r)$ are integrable. 

As an example, to illustrate how this works, consider the scalar $\mathcal{B}_1=\frac{2}{3}\,\var{\Theta} - \frac{3}{2}\,\var{\Sigma}$, for which we have
\begin{equation}
    \nonumber {\rm Var}\,\Theta = \frac{\int \mathrm{d}r \, \left(t c_1(r) + t_0 c_2(r)\right)\int \mathrm{d}r' \, \frac{\left(2tc_1(r')+t_0 c_2(r')\right)^2}{tc_1(r')+ t_0 c_2(r')} - \left(\int\mathrm{d}r \,\left(2tc_1(r)+ t_0 c_2(r)\right)\right)^2}{t^2\left(\int \mathrm{d}r \, \left(t c_1(r) + t_0 c_2(r)\right)\right)^2}\,,
\end{equation}
and
\begin{equation}
    \nonumber {\rm Var}\,\Sigma = \frac{4}{9t^2}\,\frac{\int \mathrm{d}r \, \left(t c_1(r) + t_0 c_2(r)\right)\int \mathrm{d}r' \,\frac{t_0^2 c_2(r')^2}{tc_1(r')+ t_0 c_2(r')} - \left(\int \mathrm{d}r \, t_0 c_2(r)\right)^2 }{\left(\int \mathrm{d}r \,\left(tc_1(r)+ t_0 c_2(r)\right)\right)^2}\,,
\end{equation}
where all integrals should be understood to be between $r_{\mathrm{min}}$ and $r_{\mathrm{max}}\,$. This clearly demonstrates that $\mathcal{B}_1$ vanishes, as long as all the integrals in the two expressions above are well-defined. Similar calculations show that the other $\mathcal{B}_i$ also vanish. 

This pleasing result can be understood by thinking about the averages of $c_{1}(r)$ and $c_{2}(r)$: 
\begin{equation}
\left\langle c_{1,2}\right\rangle(t) := \frac{\int\mathrm{d}r\,\left(tc_1(r)+t_0 c_2(r)\right)c_{1,2}(r)}{\int \mathrm{d}r\,\left(tc_1(r)+ t_0 c_2(r)\right)}\,,
\end{equation}
where the averages pick up a time dependence due to the presence of $t$ in the integrands. From these, we can define an effective line-element
\begin{equation}\label{eq_eff_line_element_BI}
    \mathrm{d}s^2_{\rm eff} = -\mathrm{d}t^2 + A^2(t)\,\mathrm{d}r^2 + B^2(t)\,\left(\mathrm{d}y^2 + \mathrm{d}z^2\right),
\end{equation}
where 
\begin{equation}\label{eq_Bianchi_I_Aeff}
    A(t) = \left\langle c_1\right\rangle(t) \,\left(\frac{t}{t_0}\right)^{2/3} + \left\langle c_2\right\rangle(t)\,\left(\frac{t}{t_0}\right)^{-1/3}\qquad {\rm and} \qquad B(t) = \left(\frac{t}{t_0}\right)^{2/3} \, .
\end{equation}
This means that the averaged geometry behaves like a degenerate (i.e. LRS) Bianchi type-$I$ cosmology, with directionally-dependent scale factors $A(t)$ and $B(t)\,$. As the metric in Eq. (\ref{eq_eff_line_element_BI}) is a member of the target space of solutions in our averaging formalism (i.e. it is an LRS Bianchi space-time), there is therefore no back-reaction. This result means that we can always identify a unique homogeneous model that describes the large-scale dynamics, as long as we are prepared for that model to be anisotropic\footnote{Note, however, that if one were to take the target space for one's averaging procedure to be FLRW, as in most approaches to scalar averaging in cosmology, then this effective line-element cannot be mapped exactly onto that space. Hence, performing averages that map this class of space-times onto an FLRW cosmology must necessarily involve some non-zero amount of back-reaction. This exemplifies the usefulness of our approach, which is designed specifically for understanding space-times with large-scale anisotropy. It also provides a way to understand the result that the square of the shear, $\avg{\sigma^2}$, need not be small \cite{marozzi2012late}, which would usually be interpreted as a contribution to the back-reaction scalar $\mathcal{B}$, but in the present case would be accounted for by the emergent large-scale anisotropy.}.

Finally, the quantities required to solve Eqs. (\ref{sachs_old_1})-(\ref{eq_dz_dlambda}) in the $R' = 0$ class of plane-symmetric space-times are
\begin{eqnarray*}\label{eq:H_parallel_R'=0}
    \hspace{-1cm}
    H^{\parallel} &=& \frac{2tc_1 + t_0 c_2}{3t\left(tc_1 + t_0 c_2\right)} 
    + \frac{\left(\frac{t}{t_0}\right)^{2/3} t_0^3 c_2 \left\lbrace -2k_r^2 t^2 + \left(k_y^2 + k_z^2\right)\left(t c_1 + t_0 c_2\right)^2\right\rbrace}{3k_t^2 t^3 \left(tc_1 + t_0 c_2\right)^3}\,, \\
\label{eq:ricci_R'=0}
    \hspace{-1cm} \, \Phi_{00} &=& -\frac{c_1}{3\left(tc_1 + t_0 c_2\right)}\left[\frac{k_r^2 t_0}{\left(\frac{t}{t_0}\right)^{1/3}\left(tc_1+t_0c_2\right)^2} + \frac{k_t^2 t^2 + \left(k_y^2 + k_z\right)^2 t_0^2 \left(\frac{t}{t_0}\right)^{2/3}}{t^3}\right]\,,\\
\label{eq:weyl_R'=0}
\Psi_0 &=& \frac{\left(k_y^2 + k_z^2\right) t_0 c_2 \left[k_r^2 t^2 + \left(k_t^2 \left(\frac{t}{t_0}\right)^{4/3} + \left(k_y^2 + k_z^2\right)\right)\left(tc_1 + t_0 c_2\right)^2 \right]}{3t^2 \left(\frac{t}{t_0}\right)^{4/3}\left(tc_1 + t_0c_2\right)\left[k_r^2 t^2 + \left(k_y^2 + k_z^2\right)\left(tc_1 + t_0c_2\right)^2\right]}\,,
\end{eqnarray*}
where $k_a$ is the tangent vector to the ray of light. With these quantities calculated as a function of affine distance along every null ray arriving at the observer, one can directly solve the Sachs optical equations (\ref{sachs1}) and (\ref{sachs2}).

\subsection{Dust-filled solutions with $R' \neq 0$}

Let us now consider the class of plane-symmetric dust-filled cosmologies with $R' \neq 0$. In this case, one has
\begin{equation}
    G_{tr} = \frac{2}{R}\left[R' \dot{\lambda} - \dot{R'}\right] {=} \, 0\,,
\end{equation}
where dots denote partial derivatives with respect to $t$, and where we have assumed $T_{tr}=0$ by aligning $\partial_t$ with the flow-lines of the dust. This equation is solved by $\lambda = \ln{R'} - \ln{f}$, where $f=f(r)$ is any arbitrary function of $r$\,. We can therefore write the metric as
\begin{equation}\label{eq_R'_neq_0_metric_general}
\mathrm{d}s^2 = -\mathrm{d}t^2 + \frac{R'^2(t,r)}{f^2(r)}\,\mathrm{d}r^2 + R^2(t,r)\left(\mathrm{d}y^2+\mathrm{d}z^2\right)\,,
\end{equation}
where we have assumed that the matter is dust and chosen the time coordinate to set $\nu=0$\,. 
The kinematic $1$+$1$+$2$-scalars are then given by
\begin{eqnarray}\label{eq_R'_neq_0_scalars}
\Theta &=& \frac{2\dot{R}}{R} + \frac{\dot{R}'}{R'} \qquad {\rm and} \qquad
\Sigma = \frac{2}{3}\left(\frac{\dot{R}'}{R'} - \frac{\dot{R}}{R}\right) \, ,
\end{eqnarray}
while the Ricci and Weyl scalars are
\begin{eqnarray}
\rho &=& \frac{-2Rff' - f^2 R' + \dot{R}\left(R'\dot{R} + 2R\dot{R}'\right)}{R^2 R'}
\\
\mathcal{E} &=& \frac{-Rff' + R'\left(f^2 + R\ddot{R}-\dot{R}^2\right) + R\left(\dot{R}\dot{R}' - R\ddot{R}'\right)}{3R^2 R'}\, ,
\end{eqnarray}
and $\phi = {2f}/{R}$. We note that in the $R'\neq 0$ class we have $\phi\neq 0$\,, in contrast to what happens in the case $R' = 0\,$.

Within this class, the rest of Einstein's equations are solved completely if we write the following constraint equation \cite{Stephani_2003}
\begin{equation}\label{eq_constraint_rdotsq}
    \dot{R}^2 - f^2(r) = \frac{2m(r)}{R(t,r)}\, ,
\end{equation}
where $m(r)$ is another arbitrary function, which can be related to the energy density of the dust according to
\begin{equation}\label{eq_mu_mprime}
    \rho(t,r) = \frac{2m'(r)}{R^2 R'(t,r)}\,.
\end{equation}
Eq. (\ref{eq_constraint_rdotsq}) is solved in parametric form by
\begin{eqnarray}\label{eq_R'_neq_0_solution_parametric}
    R(t,r) &=& \frac{m(r)}{f^2(r)}\left(\cosh{\eta}-1\right)\,, \\
    \nonumber t - t_0(r) &=& \frac{m(r)}{f^3(r)}\left(\sinh{\eta}-\eta\right)\,,
\end{eqnarray}
where $t_0(r)$ is a third free function, which can be thought of as setting the bang time at each value of $r$\,, such that the coordinate extent of the space-time is bounded by the curve $t = t_0(r)\,$.
A choice of the three free functions $f(r)$, $m(r)$ and $t_0(r)$ specifies a solution to Einstein's equations of the form given in Eq. (\ref{eq_R'_neq_0_metric_general}). These constitute two independent functional degrees of freedom, as there remains a freedom in re-parameterising the $r$ coordinate.

In the present case, the plane symmetry of the space-time means that calculating the Buchert averages reduces to computing a set of one-dimensional integrals of the form
\begin{equation}
\avg{S}(t) = \frac{\int_{r_{\rm min}}^{r_{\rm max}} \mathrm{d}r\, R^2(t,r) \, \left\vert \frac{R'(t,r)}{f(r)}\right\vert \, S(t,r)}{\int_{r_{\rm min}}^{r_{\rm max}} \mathrm{d}r\, R^2(t,r) \, \left\vert \frac{R'(t,r)}{f(r)}\right\vert }\: .
\end{equation}
Aside from a small number of special cases, the back-reaction scalars are generically non-zero in these solutions, as will be verified explicitly in Sections \ref{sec:farnsworth} and \ref{sec:linear}.
 
For any set of functions $f(r)$, $m(r)$ and $t_0(r)$, the quantities required to solve Eqs. (\ref{sachs_old_1})-(\ref{eq_dz_dlambda}) in the $R' \neq 0$ class are
\begin{eqnarray*}
\hspace{-0.5cm}H^{\parallel} &=& \frac{2k_r^2 f^2 R^2\left(\dot{R}' R - R' \dot{R}\right) + R'^2\left\lbrace \left(k_y^2 + k_z^2 + 2k_t^2 R^2\right)R'\dot{R}-\left(k_y^2 + k_z^2 - k_t^2 R^2\right)R\dot{R}'\right\rbrace}{3k_t^2 R^3 R'^3}\,
\\
\hspace{-0.5cm}\Phi_{00} &=& \frac{1}{2R^4 R'^3}\Big[2k_r^2 f^3 f' R^3 + \left(k_y^2 + k_z^2\right) ff'RR'^2 + f^2\left\lbrace \left(k_y^2 + k_z^2\right)R'^3 - k_r^2 R^3 \left(2\dot{R}\dot{R}' + R\ddot{R}'\right)\right\rbrace \\
\hspace{-0.5cm}&&\hspace{1.5cm}  - R'^2 \left\lbrace \left(k_y^2 + k_z^2\right)R\dot{R}\dot{R}' + R'\left(\left(k_y^2 + k_z^2\right)\dot{R}^2 + \left(k_y^2 + k_z^2 - 2k_t^2 R^2 \right)R\ddot{R}\right) - k_t^2 R^4 \ddot{R}'\right\rbrace\Big]\,,
\\
\hspace{-0.5cm}\Psi_0 &=& \frac{\left(k_y^2 + k_z^2\right)\left(k_r^2 f^2 R^2 + \left(k_y^2 + k_z^2 + k_t^2 R^2\right)R'^2 \right)\left[Rff' + R'\left(\dot{R}^2 - R\ddot{R} - f^2\right) + R\left(R\ddot{R}' - \dot{R}\dot{R}'\right)\right]}{4R^4 R' \left[k_r^2 f^2 R^2 + \left(k_y^2 + k_z^2\right)R'^2\right]}\,,
\end{eqnarray*}
where $k_a$ is again the tangent vector to a ray of light. We will use these equations to create Hubble diagrams in tilted and inhomogeneous space-times in Sections \ref{sec:farnsworth} and \ref{sec:linear}, below.

\section{An $R'=0$ Universe with Inhomogeneity}\label{sec:sinusoidal}

In this section we consider Hubble diagrams constructed in plane-symmetric dust-dominated cosmologies with $R' = 0$, in which all $\mathcal{B}_i=0$. This means that the average evolution of the cosmology is exactly equivalent to that of a Bianchi model, and the metric can be written as in Eq. (\ref{Rprime=0_metric}). We note that in such cases Buchert's back-reaction scalar $\mathcal{B}$ does not vanish \cite{Clifton_2019}, even though the scalars $\mathcal{B}_i$ from our anisotropic formalism are all zero. 

\begin{figure}[h!] 
    \centering
    \includegraphics[width=0.6\linewidth]{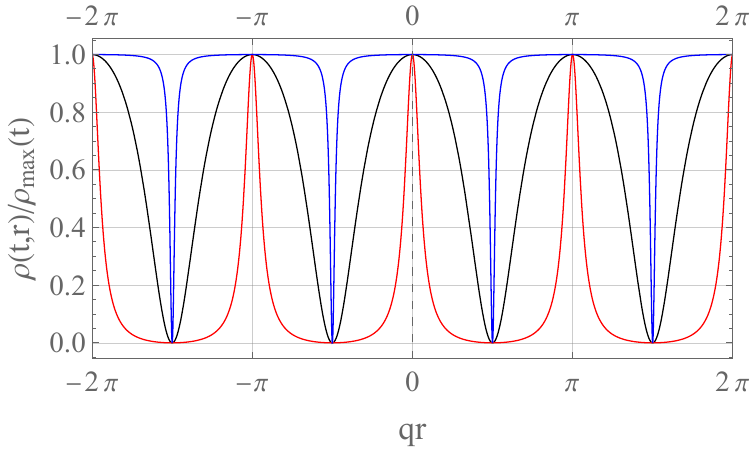}
    \caption{Energy density profile on constant-time hypersurfaces, normalised by its maximum value, in the $R'=0$ geometry from Eq. \ref{Rprime=0_metric}, with $c_1(r) = 2\cos^2{qr}$ and $c_2(r) = 2\eta\sin^2{qr}$\,. Curves correspond to $t=t_0$ (black), $t=t_0/100$ (red) and $t=100\,t_0$ (blue), and we have chosen $\eta = 1/3\,$.}
    \label{fig:1}
\end{figure}

Let us construct a space-time within this class that exhibits non-perturbative inhomogeneity in the matter distribution. This can be achieved by choosing the free functions $c_1(r)$ and $c_2(r)$ to be oscillatory, such that $c_1(r) = 2 \cos^2{qr}$ and $c_2(r) = 2\eta \sin^2{qr}$. The energy density of the dust, as measured by comoving observers, is then
\begin{equation}
\rho(t,r) = \frac{8\cos^2{qr}}{3t\left(t+\eta + \left(t-\eta\right)\cos{2qr}\right)}\ ,
\end{equation}
where for simplicity we have normalised the time coordinate so that $t_0 =1$. We see in Fig. \ref{fig:1} that for $t \ll \eta$, $\rho \longrightarrow 4\tan^2{qr}/\left(3t\eta\right)$, so the density profile is dominated by Kasner-like vacuum at early times, with small regions of very high density, whereas for $t \gg \eta$, $\rho \longrightarrow 4/\left(3t^2\right)$, as the density profile tends towards that of an homogeneous Einstein-de Sitter (EdS) universe.

Correspondingly, we have from Eq. (\ref{eq_R'=0_scalars}) that the expansion, shear and electric Weyl scalars are
\begin{eqnarray}
\Theta(t,r) &=& \frac{2t + \eta + \left(2t-\eta\right) \cos{2qr}}{t\left[t + \eta + \left(t-\eta\right)\cos{2qr}\right]}\: ,\\
\Sigma(t,r) &=& \frac{-4\eta \sin^2{qr}}{3t\left[t + \eta + \left(t-\eta\right)\cos{2qr}\right]}\: ,\\
\mathcal{E}(t,r) &=& \frac{-8\eta \sin^2{qr}}{9t^2\left[t + \eta + \left(t-\eta\right)\cos{2qr}\right]}\: = \frac{2}{3t}\,\Sigma(t,r) \, .
\end{eqnarray}
From these one may reconstruct the shear tensor as $\sigma_{ab} = \Sigma \left(3m_a m_b - n_a n_b - g_{ab}\right)/2\,$, and the electric part of the Weyl tensor as $E_{ab} = \mathcal{E} \left(3m_a m_b - n_a n_b - g_{ab}\right)/2\,$.  The expansion scalar $\Theta$ behaves in a similar fashion to the plot of $\rho$ in Fig. \ref{fig:1}, while the scalars $\mathcal{E}$ and $\Sigma$ have the opposite behaviour. At early (vacuum-dominated) times $\mathcal{E}$ and $\Sigma$ are mostly large and non-zero (as in Kasner), whereas at late (matter-dominated) times they are mostly zero (as in EdS), except for spikes in the vacuum regions. 

\subsection{Ray tracing}

We solve the geodesic equations (\ref{eq_geodesic_equation}) for a large number of observing directions $\theta_c$, and observer positions $r_0$, for observers at time $t = t_0$. Because of the plane symmetry of the space-time, the initial coordinates $y_0$ and $z_0$ are irrelevant, as is the azimuthal angle $\phi_c$ on the observer's celestial sphere (we choose $\phi_c = 5\pi/4$, for the sake of numerical simplicity). Moreover, the symmetry of the sinusoidal metric profile means that one only need consider $\theta_c$ in the range $[0,\pi/2]\,$. We can then numerically integrate the geodesic and Sachs equations.  The free parameters in the metric are set to $\eta = 1/3$ and $q = 100\,$. 
We normalise the time coordinate so that the observing time $t_{\rm obs} = t_0$ is equal to 1, and consider $100$ observers at even $r$-coordinate spacings between $r = 0$ and $r = \pi/q$\,. The angular range is split up into discrete intervals of $\Delta \,\theta_c = \pi/200\,$.

\begin{figure}[h!]
{\phantom{a}}
\hspace{-2cm}
    \begin{subfigure}{0.55\textwidth}
    \includegraphics[width=1\linewidth]{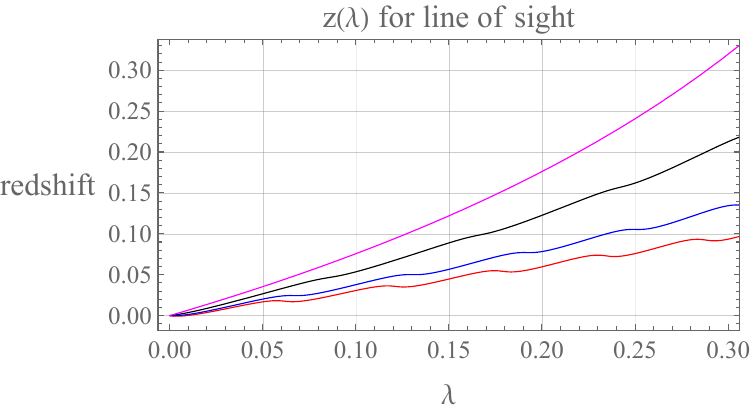}
    \caption{}
    \label{fig:z_lambda_sinusoidal}
    \end{subfigure}
    \begin{subfigure}{0.55\textwidth}
    \includegraphics[width=\linewidth]{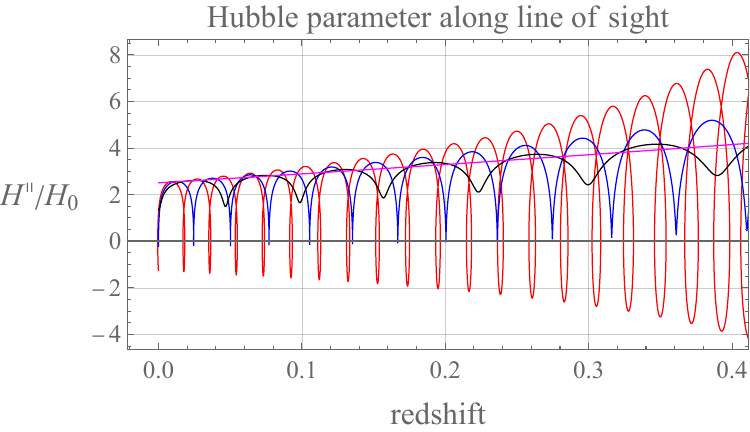} 
    \caption{}
    \label{fig:H_of_z_sinusoidal}
    \end{subfigure}
\newline
{\phantom{a}}
\hspace{-2cm}
    \begin{subfigure}{0.55\textwidth}
    \includegraphics[width=\linewidth]{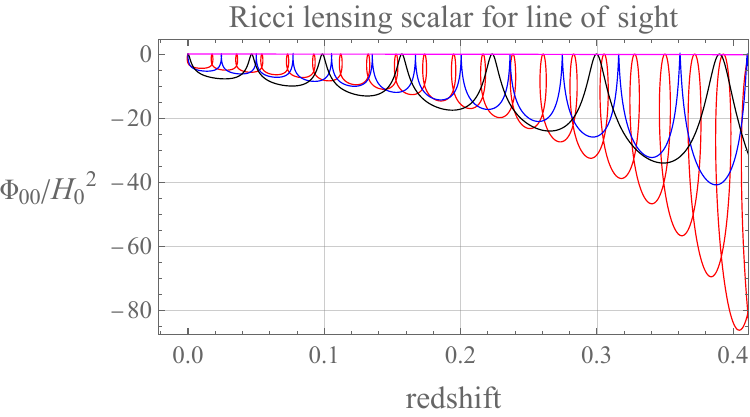} 
    \caption{}
    \label{fig:ric_sinusoidal}
    \end{subfigure}
    \begin{subfigure}{0.55\textwidth}
    \includegraphics[width=\linewidth]{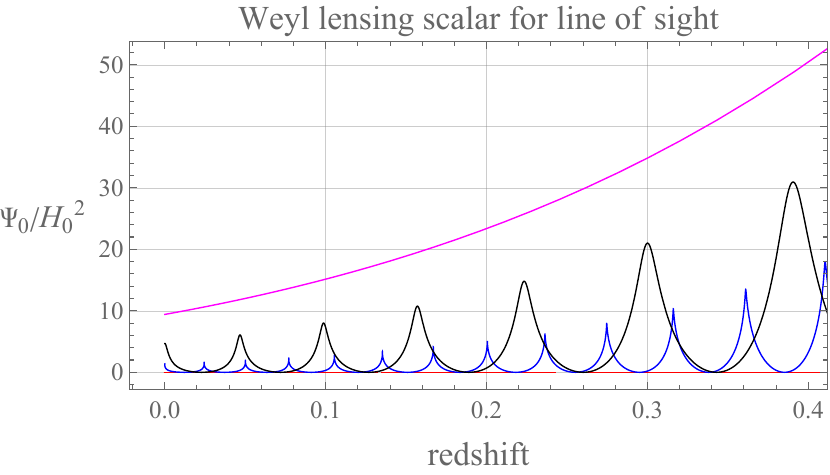} 
    \caption{}
    \label{fig:weyl_sinusoidal}
    \end{subfigure}
    \vspace{-0.4cm}
    \caption{Top left panel: photon redshift $z$ as a function of affine parameter $\lambda$, for observing angles $\theta_c = 0$ (red), $\pi/8$ (blue), $\pi/4$ (black) and $\pi/2$ (magenta) in the $R'=0$ geometry. Top right: line-of-sight Hubble parameter $H^{\parallel} = \Theta/3 + \sigma_{ab}e^a e^b$ (normalised by its monopole $H_0$ at $z = 0$). Bottom left and right: Ricci and Weyl lensing scalars $\Phi_{00} = -1/2 R_{ab}k^a k^b$ and $\Psi_0 = C_{abcd} s^a k^b s^c k^d$ as functions of redshift for the same observer (both made dimensionless by normalising with respect to $H_0^2$).}
    \label{fig:sinusoidal_raytracing}
\end{figure}

Some key results of carrying out the ray tracing are shown in Fig. \ref{fig:sinusoidal_raytracing}. In these plots, we have considered an observer at time $t_0$ who is placed at a point $r_{\rm obs} = \pi/2q$, which corresponds to the centre of an underdensity, i.e. $\rho(t_0, r_{\rm obs}) = 0$.
For each initial direction $\theta_c$, we can first calculate the redshift $z$ as a function of the affine parameter in the geodesic equation (\ref{eq_geodesic_equation}). For small $\theta_c$ (e.g. the red and blue curves in Fig. \ref{fig:z_lambda_sinusoidal}), the bumpy nature of the function $z(\lambda)$ indicates the strong oscillatory inhomogeneities in the metric as light rays propagate in that direction. 
For those directions, the function $z(\lambda)$ is not monotonically increasing. When a future-directed null ray passes through the vacuum-dominated regions, it can gain energy as it moves forward in time (and vice versa for past-directed null rays), if it is directed along or close to the spatially contracting symmetry axis of the rotationally symmetric Kasner-like geometry.
For large $\theta_c$ (e.g. the black and pink curves), $z(\lambda)$ is much smoother, because those observers are looking in directions along which the space-time is much closer to homogeneous.

With $z(\lambda)$ obtained, one can then calculate any quantity as a function of the observable redshift, rather than the non-observable $\lambda$. In Fig. \ref{fig:H_of_z_sinusoidal} we show the expansion rate $H^{\parallel}$ parallel to the corresponding null geodesic, as a function of the redshift along that curve. As the light ray passes through the underdense regions, the function $H^{\parallel}$ dips and can become negative if the effect of the inhomogeneity is sufficiently strong, as can happen for small $\theta_c$. 
The loops in the $H^{\parallel}$ curve for $\theta_c = 0$ reflect the non-monotonicity of $z(\lambda)$, wherein a past-directed null ray moving from an overdense to an underdense region initially has $H^{\parallel}$ decreasing but $\mathrm{d}z/\mathrm{d}\lambda$ remaining positive. This produces the leading, right-hand side of each loop. Then as the light ray is moving towards the centre of the underdensity, $\mathrm{d}z/\mathrm{d}\lambda$ becomes negative as $H^{\parallel}$ crosses zero, as per Eq. (\ref{eq_dz_dlambda}). 
After the null ray passes the centre of the underdensity, $H^{\parallel}$ begins to increase again. Finally $H^{\parallel}$ passes back through zero, and therefore the redshift begins to increase again. For null rays travelling in directions where $H^{\parallel}$ is always positive, these loops do not exist. Moreover, as seen in the pink curve in Fig. \ref{fig:H_of_z_sinusoidal}, for $\theta_c = \pi/2$ the geometry that the ray moves through appears spatially homogeneous, and $H^{\parallel}$ increases monotonically.

Because we are displaying the results for observers situated within a vacuum region, the expansion rate $H^{\parallel}_0$ at redshift zero is large and negative for $\theta_c = 0$, with $\vert H_0^{\parallel} \vert$ similar to the magnitude of the all-sky average of $H_0$. The expansion rate of space along the line of sight then monotonically increases as a function of $\theta_c$, to over twice the all-sky average at $\theta_c = \pi/2\,$. This is indicative of the very strong anisotropy in the space-time. One sees a similar set of effects for the Ricci and Weyl curvature terms $\Phi_{00}$ and $\Psi_0$, in Figs. \ref{fig:ric_sinusoidal} and \ref{fig:weyl_sinusoidal}, respectively.  The scalar $\Psi_0$ is small in matter-dominated regions, where the EdS-like dynamics dominate, and rises sharply in the Kasner-like underdensities. For null rays with $\theta_c$ near $\pi/2$, $\Psi_0$ increases monotonically, and $\Psi_0 \gg \vert \Phi_{00} \vert$ at all redshifts. The first of these facts is explained, like for the $H^{\parallel}$ curve in Fig. \ref{fig:H_of_z_sinusoidal}, by the apparent homogeneity along that line of sight. For the latter fact, we recall that $\Phi_{00} = -R_{ab}k^a k^b/2 = -T_{ab}k^a k^b/2$. Therefore, for a light ray that always resides in regions of zero (or very low) matter density, the energy-momentum tensor remains zero (or near zero), and so the space-time curvature is dominated instead by the free gravitational field encoded in the Weyl curvature.

For null rays with $\theta_c = 0$, $\Phi_{00}$ is large (and negative, as it must be due to the null energy condition) at most points where the density is sufficiently large that the dynamics are EdS-like, and then drops sharply to zero in vacuum/near-vacuum regions. 
Note, however, that the Weyl scalar $\Psi_0$ remains zero throughout, because a null congruence that travels directly along the rotational symmetry axis of the space-time cannot be sheared (it is a principal null direction). The loops in Fig. \ref{fig:ric_sinusoidal} have the same origin as those in Fig. \ref{fig:H_of_z_sinusoidal}. They are just about visible in the blue curve ($\theta_c = \pi/8$) in Fig. \ref{fig:weyl_sinusoidal}, for which the Weyl term is non-zero, but is strongly suppressed relative to the Ricci curvature.

With past-directed solutions to the geodesic equation (\ref{eq_geodesic_equation}) known, we can now solve the Sachs equations (\ref{sachs1}) and (\ref{sachs2}) to obtain the angular diameter distance $d_A(z)$, and hence the luminosity distance, along individual lines of sight. This will be done for each possible $\theta_c$, at multiple observer locations $r_{\rm obs}$ on the $t = t_0$ constant-time hypersurface. 

\subsection{Hubble diagrams}\label{subsec:sinusoidal_hubble}

A typical approach used in cosmology is to consider the Hubble diagrams that would result in some average cosmology, which is expected to reproduce the large-scale properties of the Universe. This process is left somewhat implicit in the standard FLRW framework, but is done explicitly in the case of the Buchert averaging procedure. 
In doing this, one obtains a ``Hubble diagram of the average''. As we are currently dealing with an $R' = 0$ metric, it follows that for any choice of averaging domain the back-reaction scalars will vanish, and the large-scale model is obtained by simply replacing the sinusoidal functions $c_1(r)$ and $c_2(r)$ by their average values. 
In the present case, the average model is therefore described by a Bianchi type-$I$ metric (\ref{eq_Bianchi_I}), with scale factors $A(t) = t^{2/3} + \eta t^{-1/3}$ and $B(t) = t^{2/3}\,$. We can now perform ray tracing in that averaged model universe, and compare the results to those of the inhomogeneous model that existed before averaging. 
The averaged model in this case is homogeneous but anisotropic, meaning that the spatial location of the observer is irrelevant, but that we still expect angular variation in the Hubble diagram. 
By integrating the Sachs equations using the averaged Bianchi-$I$ metric, we finally calculate $d_L^{\rm model}(z)$ in different directions on the observer's sky. This will let us determine how accurately that homogeneous averaged model fits the distance-redshift relation for observers in the actual inhomogeneous space-time.

\begin{figure}[h!]
{\phantom{a}}
\hspace{-1.7cm}
    \begin{subfigure}{0.50\textwidth}
    \includegraphics[width=\linewidth]{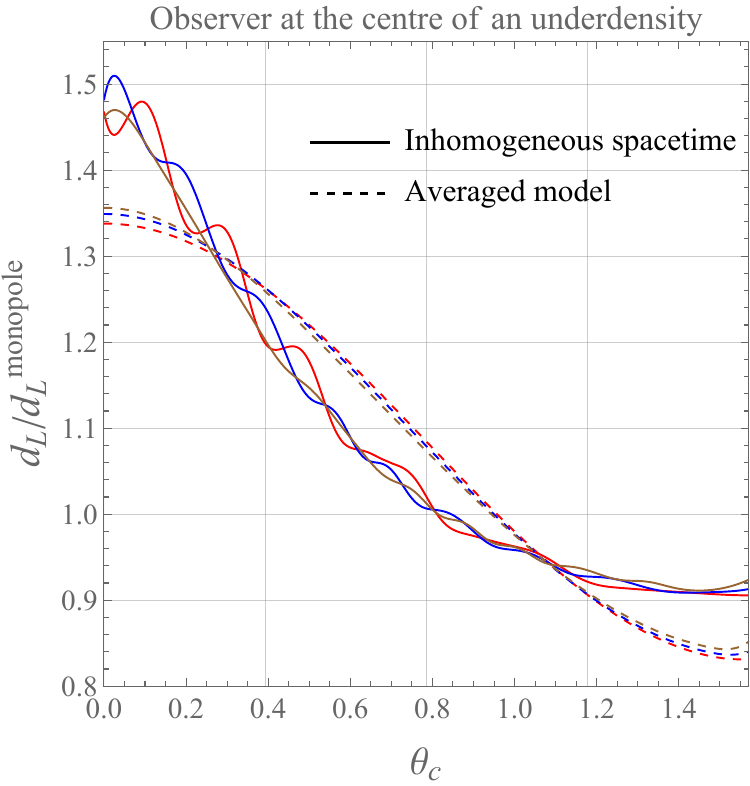} 
    \caption{}
    \label{fig:dL_over_monopole_sinusoidal}
    \end{subfigure}
 \begin{subfigure}{0.52\textwidth}
    \includegraphics[width=\linewidth]{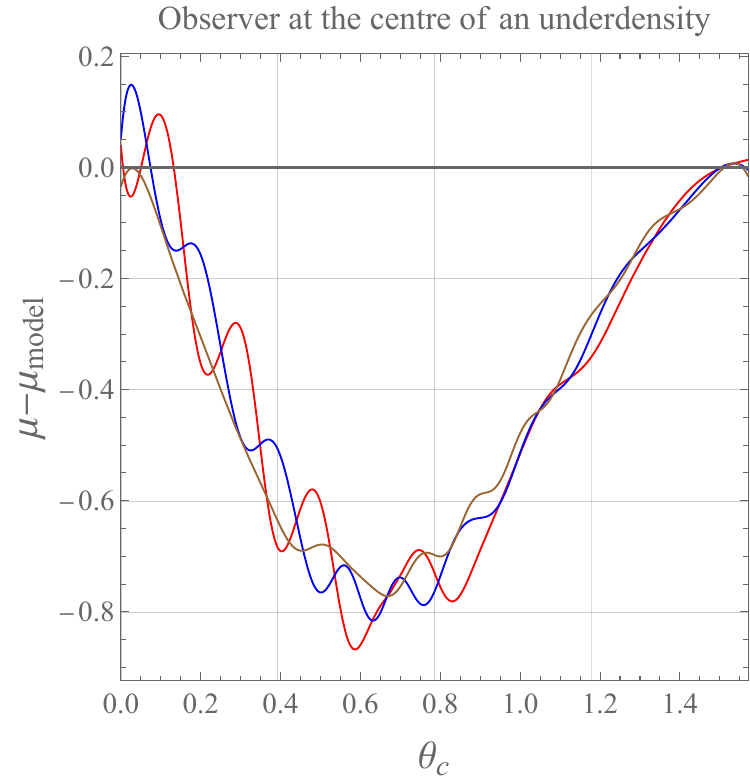} 
    \caption{}
    \label{fig:sinusoidal_model_mu-mu_model_theta}
    \end{subfigure}
\caption{Left: luminosity distance $d_L$ at a given redshift as a function of observing angle $\theta_c$, relative to the monopole, for an observer at the centre of an underdensity in the $R'=0$ geometry (as per Eq. (\ref{Rprime=0_metric}), with $c_1(r) = 2\cos^2{qr}$ and $c_2(r) = 2\eta\sin^2{qr}$\,, where $\eta = 1/3$). 
The curves are for $z = 0.1$ (red), $z = 0.2$ (blue), and $z = 0.3$ (brown). The solid lines are obtained by performing ray tracing in the inhomogeneous space-time, and the dashed lines come from ray-tracing in the averaged model. Right: difference between distance modulus $\mu$ obtained by from ray-tracing in the inhomogeneous space-time and averaged model, as a function of $\theta_c$. Curves are as given in the left-hand plot.} 
\end{figure}

\begin{figure}[b!]
\phantom{a} \hspace{-1.5cm}
    \begin{subfigure}{0.5\textwidth}
    \includegraphics[width=\linewidth]{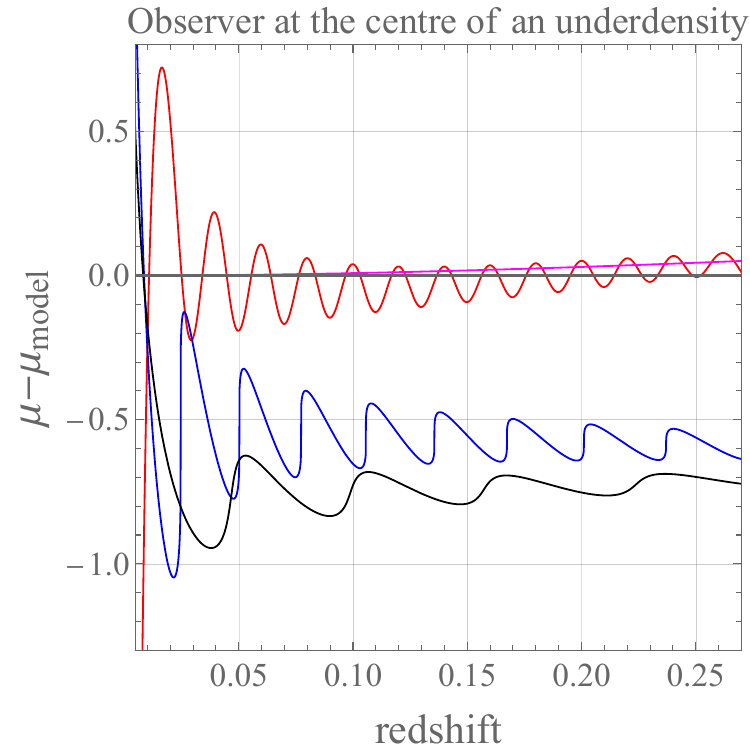} 
    \caption{}
    \label{fig:sinusoidal_magnitude_vs_model}
    \end{subfigure} \hspace{1cm}
    \begin{subfigure}{0.6\textwidth}
    \includegraphics[width=\linewidth]{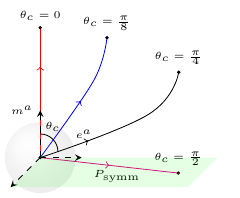}
    \vspace{-0.2cm}
    \caption{}
    \label{fig_ray_tracing_picture}
    \end{subfigure}
    \caption{Left: difference in distance modulus from the averaged model, $\mu - \mu_{\mathrm{model}}$, as a function of redshift, shown for an observer at the centre of an underdense region in the $R'=0$ geometry. The curves are for $\theta_c = 0$ (red), $\theta_c = \pi/8$ (blue), $\theta_c = \pi/4$ (black), and $\theta_c = \pi/2$ (magenta). Right: An illustration of the ray tracing procedure used to calculate the sky map of $d_L(z)$ for each observer. Past-directed null geodesics emanating from the observer are distinguished by their value of $\theta_c$, as in the left-hand plot. The space-like vector $m^a$ is indicated as being normal to the planes of symmetry $P_{\rm symm}\,$, which here are in the horizontal plane. 
    }
\end{figure}

\begin{figure}[h!]
\phantom{a} \hspace{-1.5cm}
    \begin{subfigure}{0.52\textwidth}
    \includegraphics[width=\linewidth]{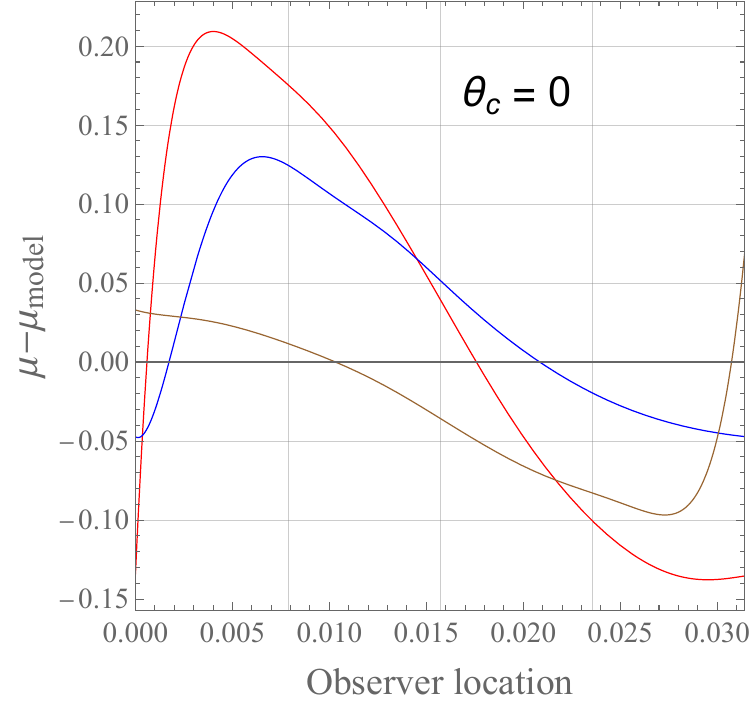} 
    \end{subfigure}
    \begin{subfigure}{0.52\textwidth}
        \includegraphics[width=\linewidth]{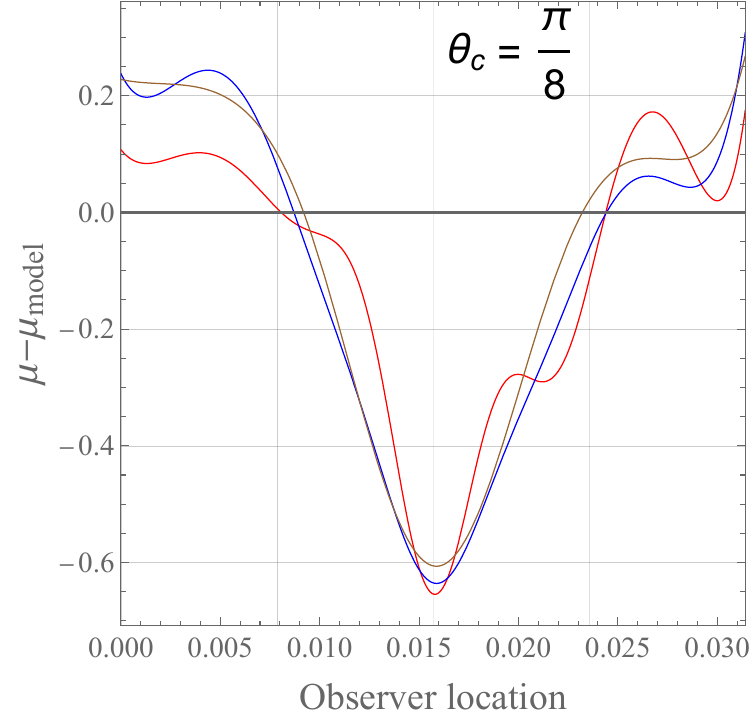} 
    \end{subfigure}
        \newline
{\phantom{a}}
\hspace{-1.5cm}
        \begin{subfigure}{0.52\textwidth}
        \includegraphics[width=\linewidth]{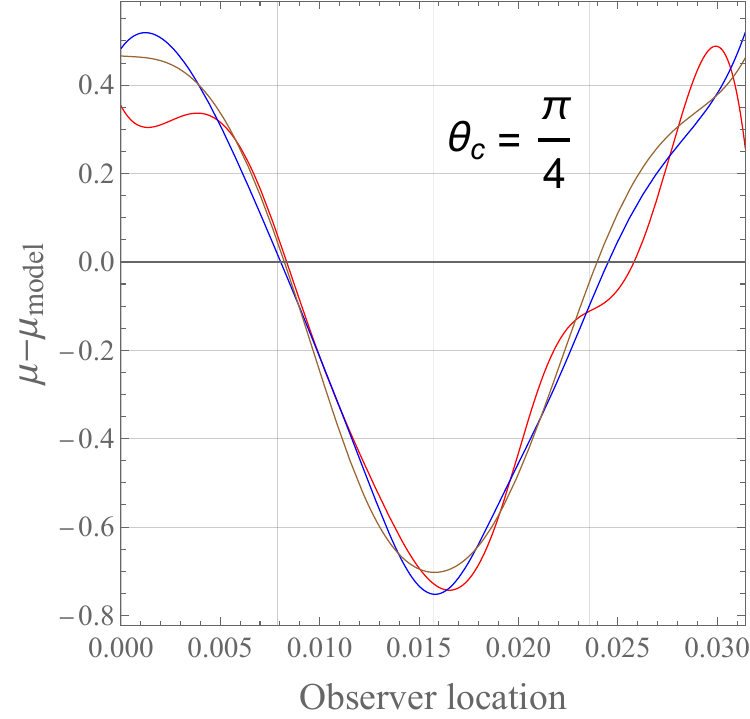} 
    \end{subfigure}
        \begin{subfigure}{0.52\textwidth}
        \includegraphics[width=\linewidth]{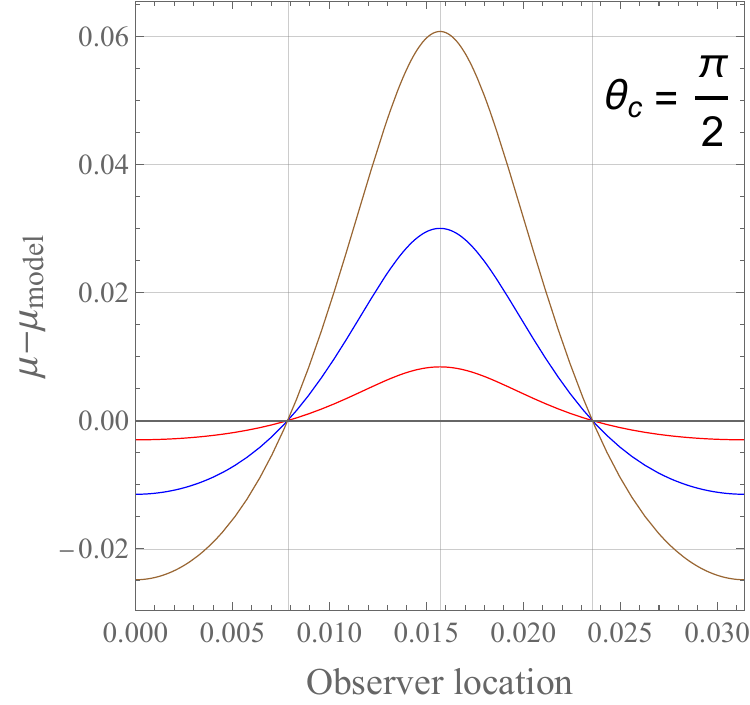} 
    \end{subfigure}
    \caption{$\mu - \mu_{\mathrm{model}}$ for a given line of sight $\theta_c$, as a function of observer location within a single averaging domain in the $R'=0$ geometry. The left and right edges of the plot correspond to overdense regions, whereas the middle corresponds to an underdensity. The  curves are for $z = 0.1$ (red), $z = 0.2$ (blue), and $z = 0.3$ (brown).}
    \label{fig:sinusoidal_mu-mu_model_vs_observer_location}
    \end{figure}

First, we consider the angular variation of $d_L$, at a constant specified redshift. Such a variation is entirely absent in an isotropic universe, but is non-trivial, and dependent on the observer location, in an inhomogeneous space-time. Our results are shown in Fig. \ref{fig:dL_over_monopole_sinusoidal}. 
The function $d_L(\theta_c)$ is rather complicated in the inhomogeneous space-time, as each light ray will have passed through a continually oscillating geometry, leading to an oscillatory pattern within the function, arising from the $\cos^2{qr}$ and $\sin^2{qr}$ terms in the metric.
That pattern is itself contained in a quadrupole envelope, which, rather than coming from any oscillations, is due to the overall discrepancy between the expansion rates in the $r$ coordinate direction and each of the $y$ and $z$ directions, producing a shear $\sigma_{ab}$ which is naturally quadrupolar and affects the Hubble diagram through e.g. the line-of-sight Hubble parameter $H_0^{\parallel}$ at $\mathcal{O}(z)\,$, the line-of-sight deceleration parameter $q_0^{\parallel}$ at $\mathcal{O}(z^2)\,$, and so on.
Thus, the envelope is a signature of the global anisotropy that arises due to the presence of Kasner-like regions, which are contracting along $\theta_c = 0$ and therefore reducing redshifts for a given luminosity distance in that direction. Therefore, to reach the target redshift requires the past-directed null geodesic to travel further, meaning that $d_L$ exceeds the all-sky average (i.e. the monopole). Conversely, observing at $\theta_c = \pi/2$ means that one sees the lowest possible $d_L$ for that redshift, as there is only EdS-like expansion along that line of sight. The averaged Bianchi-{\it I} model captures something of the quadrupolar envelope, which in that context can be interpreted entirely in terms of the shear anisotropy $\sigma_{ab}$ of the Bianchi {\it I} spacetime, but loses all information about any higher-order multipoles (i.e. multipoles with $l \geq 2$).
For observers residing in underdense, roughly Kasner-like, regions, $d_L$ is over-estimated at the axes and under-estimated in between, because the local geometry is less isotropic than the average. The extent of the average model's success (or lack thereof) in predicting the Hubble diagrams of an individual observer is displayed in Fig. \ref{fig:sinusoidal_model_mu-mu_model_theta}, wherein we see that the angular variation of $\mu - \mu_{\rm model}$ is roughly maintained as $z$ increases. 

Fig. \ref{fig:sinusoidal_magnitude_vs_model} shows the performance of the averaged homogeneous model as a function of redshift, for the same observer. The magnitude difference can be very large at low redshifts, as one expects: on very small scales, and in the presence of very large inhomogeneities, observations in one's immediate vicinity do not produce an accurate Hubble diagram of the universe at large. However, as null congruences travel through many cycles of the oscillatory geometry, $d_L$ becomes much closer to the average. This is seen most clearly in the red curve, where observing along the axis of inhomogeneity $\theta_c = 0$ means that the light ray ultimately samples equal numbers of overdensities and underdensities. It therefore converges, with an oscillatory pattern, towards the homogeneous model obtained through our averaging procedure, as it corresponds to one of the axes of the quadrupole in $d_L$ for that model. Similarly, for the pink curve, which corresponds to $\theta_c = \pi/2$, the quadrupolar nature of the average model means that $d_L$ is accurately reproduced, as also shown by the return of $\mu - \mu_{\rm model}$ to zero as a function of $\theta_c$ in Fig. \ref{fig:sinusoidal_model_mu-mu_model_theta}. For observing angles $\theta_c$ that are off the quadrupole axes, the curve $\mu(z) - \mu_{\rm model}(z)$ does not always converge back to zero, but to a constant offset value. This is to be expected for single observers, as they cannot be said to be measuring a fair sample of the universe in all directions.

\begin{figure}[b!]
\phantom{a}
\hspace{-1.7cm}
    \begin{subfigure}{0.55\textwidth}
    \includegraphics[width=\linewidth]{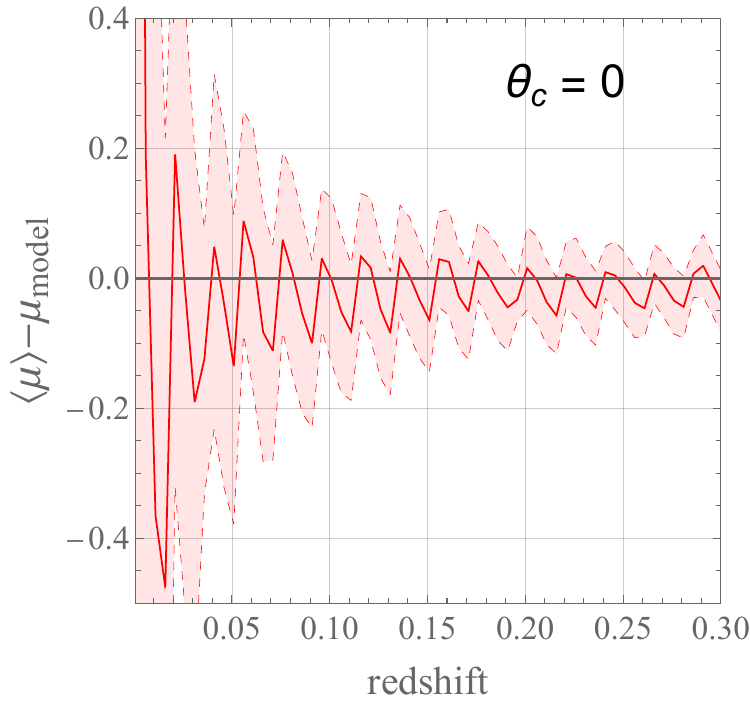} 
        \caption{}
    \end{subfigure}
    \begin{subfigure}{0.55\textwidth}
        \includegraphics[width=\linewidth]{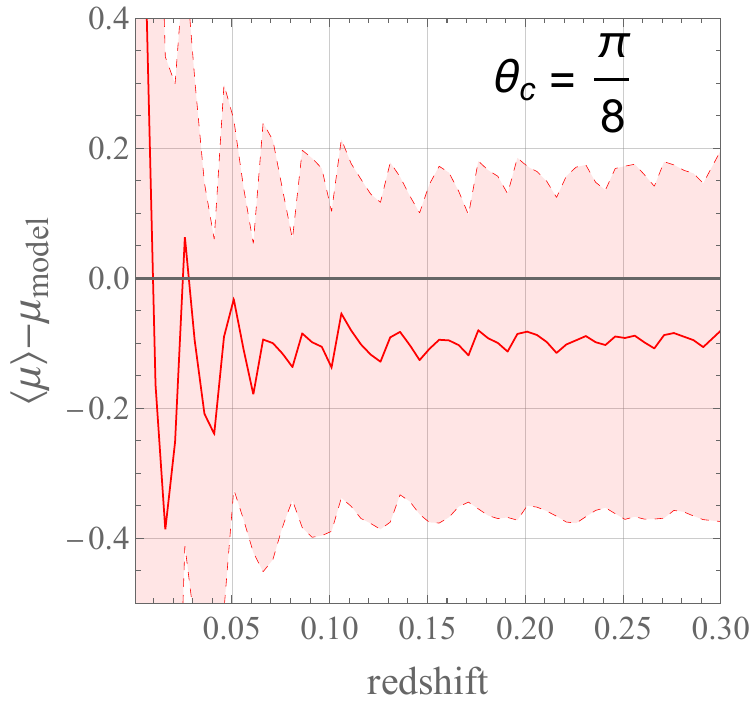} 
            \caption{}
    \end{subfigure}
    \newline
    \phantom{a}
\hspace{-1.7cm}
        \begin{subfigure}{0.55\textwidth}
        \includegraphics[width=\linewidth]{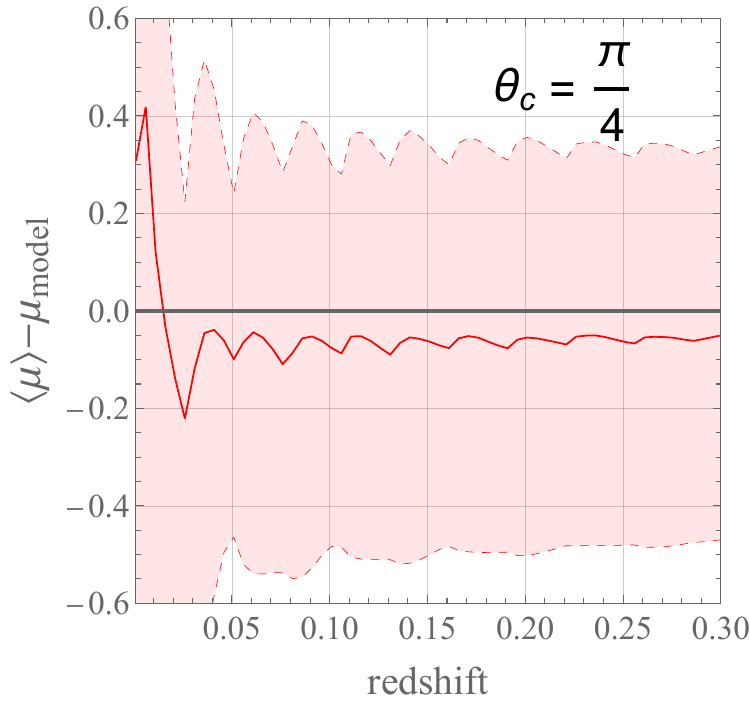} 
            \caption{}
    \end{subfigure}
        \begin{subfigure}{0.55\textwidth}
        \includegraphics[width=\linewidth]{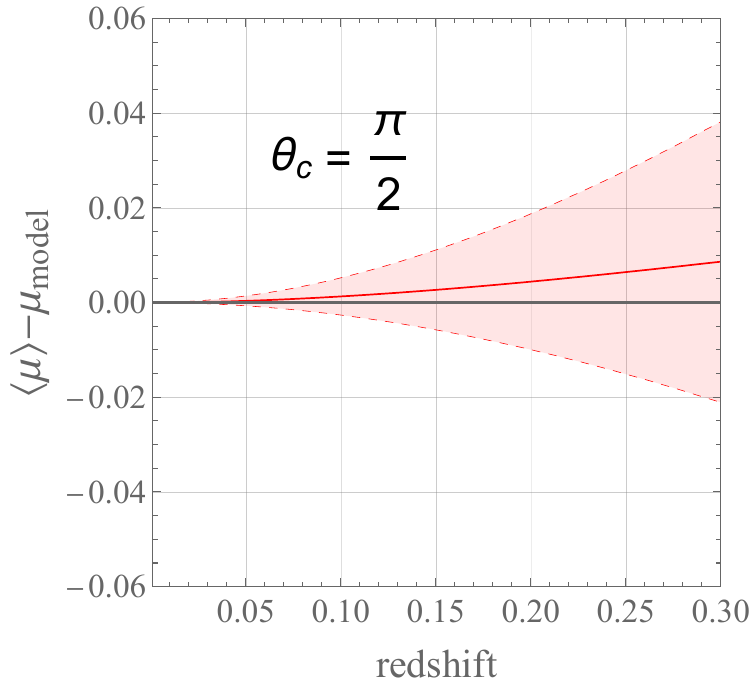} 
           \caption{}
    \end{subfigure}
    \caption{Averaged distance moduli, $\avg{\mu} - \mu_{\mathrm{model}}$, as a function of redshift, in the geometry with $R'=0$. These averages are obtained from the mean of $100$ observers evenly spaced in coordinate location $r$ throughout the averaging domain. The shaded red regions indicate the $1\sigma$ confidence intervals. Results are displayed for sets of observers all viewing in the directions $\theta_c = \left\lbrace 0, \pi/8, \pi/4, \pi/2\right\rbrace$.}
    \label{fig:sinusoidal_mean_mu_vs_model}
    \end{figure}

While any averaged model could not be expected to reproduce the Hubble diagrams of every individual observer, we might expect it to produce a good representation of the average Hubble diagram that would be obtained by averaging the results from many different observers. To investigate this possibility, we pick an observing direction, $\theta_c$, and consider the $d_L(z)$ from $100$ different observers at many different $r_{\rm obs}$ across an averaging domain, separated from one another by equal intervals of the $r$ coordinate. Comparing their measurements gives rise to a situation as in Fig. \ref{fig:sinusoidal_mu-mu_model_vs_observer_location}, where we consider the difference $\mu - \mu_{\rm model}$ for a given $\theta_c$ and $z\,$, as a function of the observer's location within the averaging domain.

By choosing a large range of redshift intervals, we can then calculate the mean and variance of $d_L$ for each point in the discretised parameter space spanned by $\left(r_{\rm obs}, \theta_c\right)$.
The averaged distance modulus $\avg{\mu}$ that results is displayed in Fig. \ref{fig:sinusoidal_mean_mu_vs_model}, where we have calculated $\avg{\mu} - \mu_{\rm model}$ for a variety of directions in the redshift range $z \in \left[0, 0.30\right)\,$. Although at low redshift the homogeneous large-scale averaged model describes the distance modulus poorly, the curve $\avg{\mu}(z)$ rapidly converges to $\mu_{\rm model}(z)$. This convergence is particularly strong for a collection of observers viewing along the axis of inhomogeneity, $\theta_c = 0$, but even for the collection of observers who are off-axis the average distance modulus can be seen to settle down to have only a small offset from the averaged model (comfortably within one standard deviation). This result reflects the capacity of our formalism to account for large-scale anisotropy.

\section{An $R' \neq 0$ Universe with Tilt}\label{sec:farnsworth}

\begin{figure}[t!]
\phantom{a}\hspace{-1.4cm}
    \begin{subfigure}{0.64\textwidth}
    \includegraphics[width=0.8\linewidth]{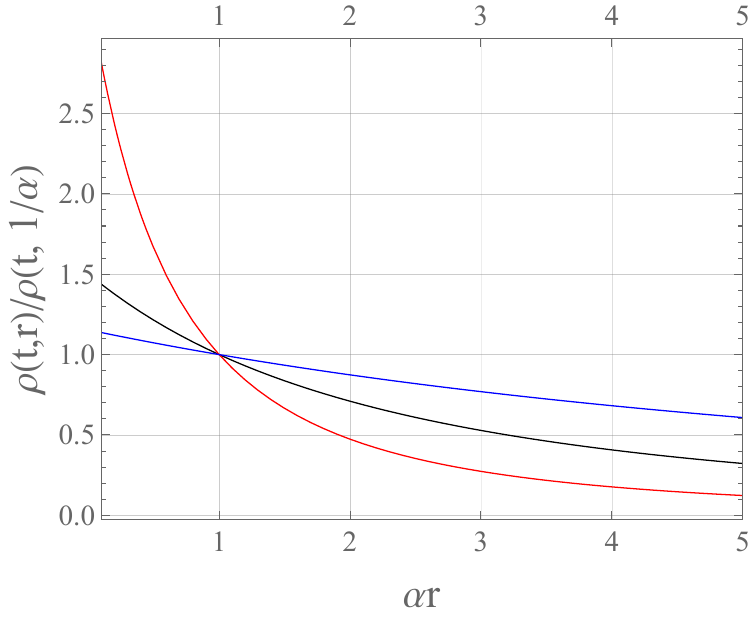}
    \caption{}
    \label{fig:rho_farnsworth}
    \end{subfigure}
    \hspace{-2cm}
    \begin{subfigure}{0.57\textwidth}
    \includegraphics[width=\linewidth]{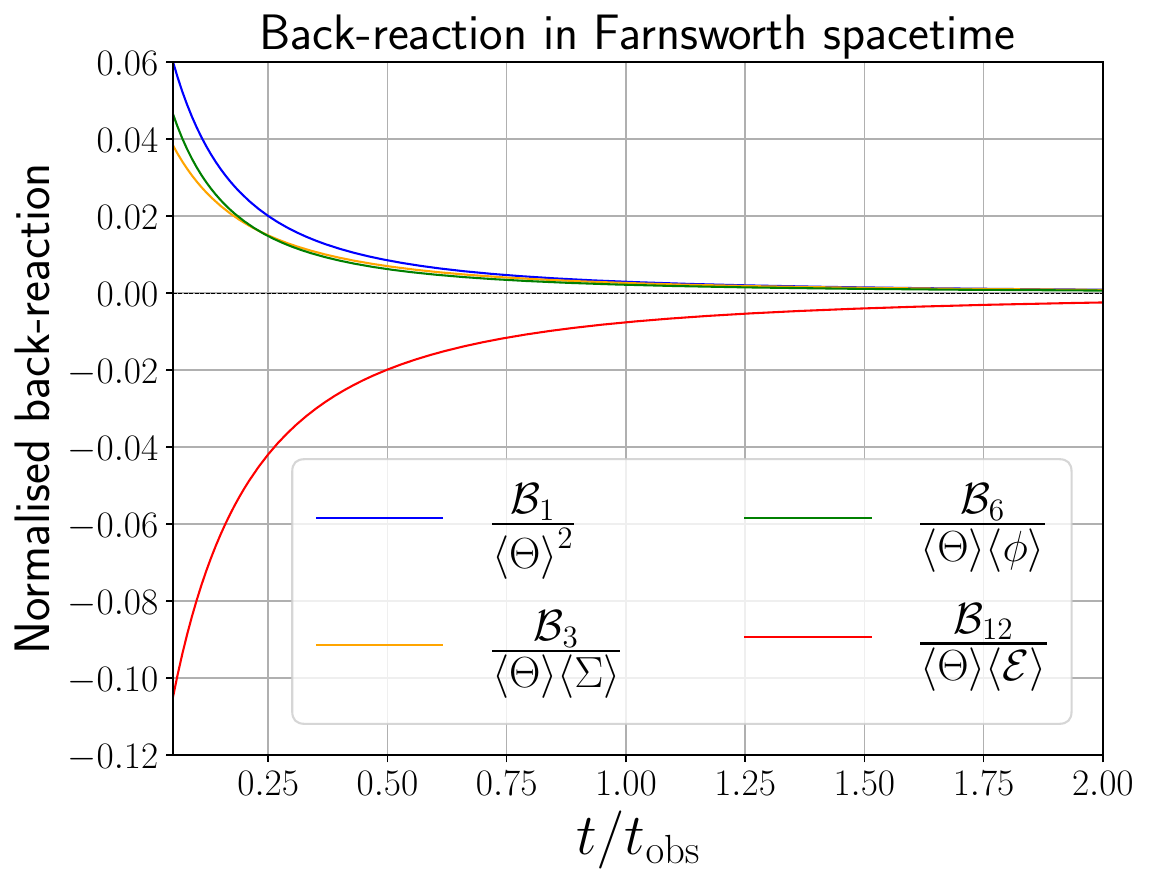}
    \caption{}
    \label{fig:backreaction_Farnsworth}
    \end{subfigure}
    \caption{Left panel: energy density profile on constant-time surfaces of the Farnsworth cosmologies, normalised by its value at $r=\alpha^{-1}$. The curves shows $\rho(r)$ at the observing time $t=t_{\rm obs}$ (black), $t=t_{\rm obs}/3$ (red), and $t=3 t_{\rm obs}$ (blue). Right panel: all non-vanishing back-reaction scalars in constant-time hypersurfaces of the Farnsworth geometry (\ref{eq_Farnsworth_definition}). The scalars $\mathcal{B}_i$ have all been made dimensionless, by normalising them with respect to the largest term from the equation in which they appear.}
    \label{fig:farnsworth_scalars}
\end{figure}

While the back-reaction scalars, $\mathcal{B}_i$, vanish in the $R' = 0$ cosmologies, the situation with $R' \neq 0$ cosmologies is more complicated. For example, the presence of a non-zero $R'$ means that the dust that sources the space-time curvature does not need to be moving along integral curves of $n^a$. This is true even if the space-time is homogeneous, and constitutes a class of anisotropic cosmological models that is usually referred to as being ``tilted''\footnote{For an introduction to tilted homogeneous cosmologies, see Ref. \cite{King_1973}.}. As in Ref. \cite{Anton_2023}, we wish to study this possibility using the anisotropic cosmologies of Farnsworth \cite{Farnsworth_1967}.

\begin{figure}[t!]
\phantom{a}\hspace{-2.4cm}
    \includegraphics[width=1.2 \linewidth]{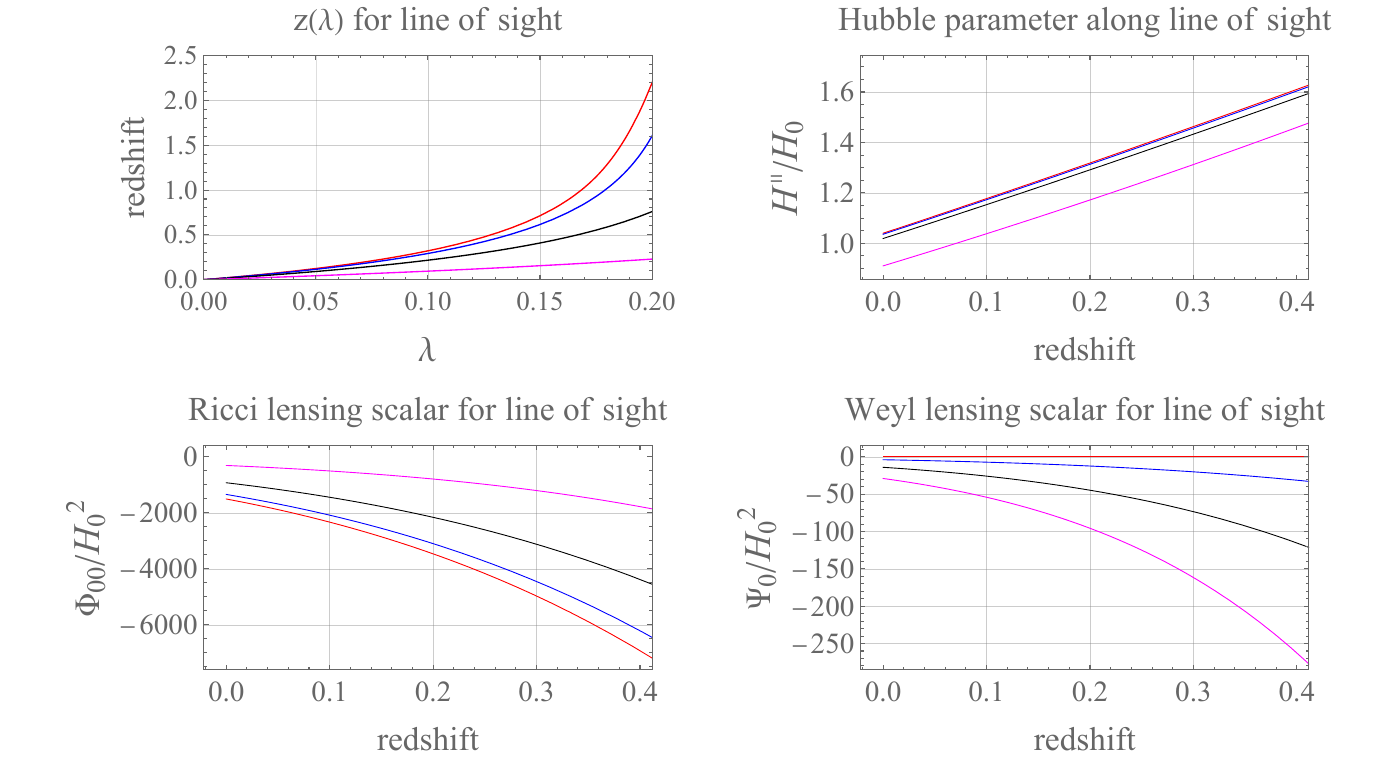}
    \caption{Top-left: redshift as a function of affine distance $\lambda$ in the Farnsworth cosmology, for observing angles $\theta_c = 0$ (red), $\pi/8$ (blue), $\pi/4$ (black) and $\pi/2$ (magenta). Top-right: line-of-sight Hubble parameter $H^{\parallel}$ as a function of redshift. Bottom-left and right: Ricci and Weyl lensing scalars $\Phi_{00}$ and $\Psi_0$, for the same observing directions.}
    \label{fig:farnsworth_raytracing}
\end{figure}

Farnsworth's cosmologies are exact homogeneous solutions of Einstein's equations of Bianchi type-$V$. They are locally rotationally symmetric, but tilted, and are given by the metric functions from Eqs. (\ref{eq_R'_neq_0_metric_general}) and (\ref{eq_R'_neq_0_solution_parametric}) as
\begin{eqnarray}\label{eq_Farnsworth_definition}
    m(r) &=& \frac{Wk^3}{2}\,e^{-3\alpha r}\,,\qquad
    f(r) = k\, e^{-\alpha r}\,,\qquad {\rm and} \qquad
    t_0(r) = -C\,r \,,
\end{eqnarray}
where $W$, $k$, $\alpha$ and $C$ are constants, with the latter determining the tilt velocity of the matter with respect to the homogeneous surfaces, which are level surfaces of the function $t + Cr$ \cite{Farnsworth_1967}.  This means that a set of observers comoving with the matter flow should measure global inhomogeneity in the hypersurface that is spanned by their rest spaces, so the ``homogeneous'' cosmological model that they would construct by averaging over domains of those hypersurfaces would appear to exhibit back-reaction, and the Hubble diagram they would infer within that model may well be a poor fit to observations of distance measures. The exception to this is the special case $C = 0$\,, in which case the solution (\ref{eq_Farnsworth_definition}) is just an FLRW geometry with negative spatial curvature.

Fig. \ref{fig:farnsworth_scalars} shows that observers comoving with the dust would have orthogonal rest spaces that are inhomogeneous, provided $C \neq 0$. This effect is entirely due to the tilt. We now wish to carry out numerical integrations for rays of light in this space-time, for which we make the following choices for parameter values: $\alpha=1$ for the characteristic length scale, and $k = 5$, $W = 125$ and $C = 2$ for the curvature, density, and tilt parameters, respectively (as in Ref. \cite{Anton_2023}). Because of the nature of the apparent inhomogeneity induced by the tilt, the rest spaces of the observers in this case are not statistically homogeneous. This means that there is no natural homogeneity scale that can be used to define our averaging domain, $\mathcal{D}$. We therefore choose to average between $\left\{r_{\rm min}, r_{\rm max}\right\} = \left\{\alpha^{-1}, 3 \alpha^{-1} \right\}\,$, which in the absence of an homogeneity scale is made purely out of computational convenience. We find that back-reaction scalars in this case can be non-zero, with relative sizes of up to $10\%\,$, though they decay at late times (as shown in Fig. \ref{fig:backreaction_Farnsworth}). 

\begin{figure}[t!]
    \phantom{a}\hspace{-1.5cm}
    \begin{subfigure}{0.53\textwidth}
    \includegraphics[width=\linewidth]{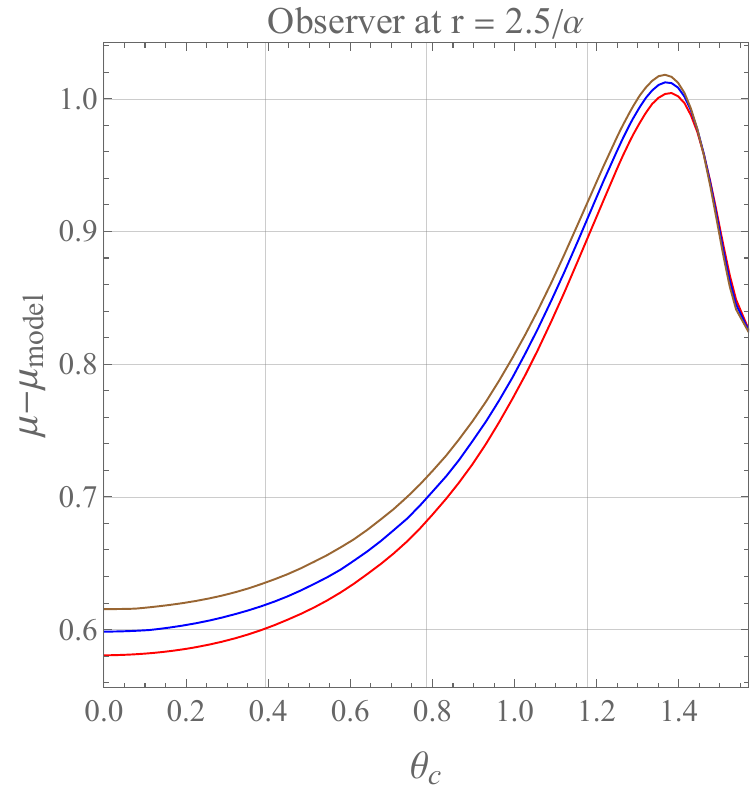} 
    \caption{}
    \label{fig:Farnsworth_mu-mu_model_theta}  
    \end{subfigure}
    \begin{subfigure}{0.55\textwidth}
    \includegraphics[width=\linewidth]{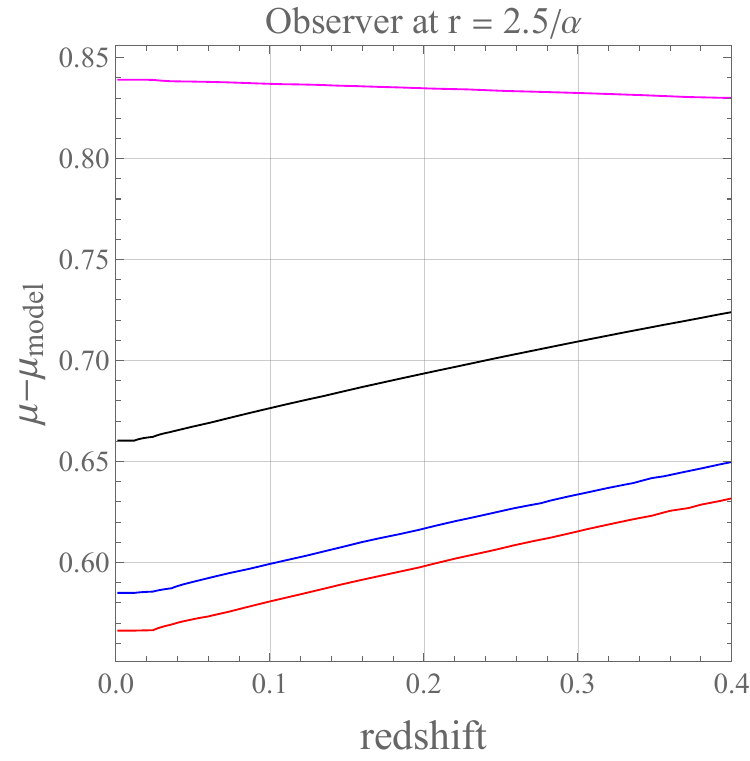} 
    \caption{}
    \label{fig:Farnsworth_magnitude_vs_model}
    \end{subfigure}
\caption{Left: $\mu - \mu_{\rm model}$ as a function of $\theta_c$, for at observer at $r_{\rm obs} = 2.5/\alpha$ in the Farnsworth cosmology. Curves are for redshifts $0.1$ (red), $0.2$ (blue) and $0.3$ (brown). The solid lines are obtained by performing ray tracing in the tilted space-time, while the dashed lines come from ray tracing in the averaged LRS Bianchi type-$V$ model.
Right: $\mu - \mu_{\rm model}$ as function of redshift, for the observer at $r = 2.5/\alpha$ considered in the last set of plots. Curves are for observing angles $\theta_c = 0$ (red), $\pi/8$ (blue), $\pi/4$ (black) and $\pi/2$ (magenta). Again, $\mu_{\rm model}$ refers to a Bianchi type-$V$ model.}
\end{figure}

\subsection{Ray tracing}

We now calculate the paths of null geodesics arriving at observers on a constant-time hypersurface, by repeatedly solving the geodesic equation in this space-time. Once again, the plane symmetry of the space-time restricts the initial conditions we need to vary to $r_{\rm obs}$ and $\theta_c$. We note that although the space-time has homogeneous surfaces, these are not the constant-time surfaces of observers comoving with the dust, and so we would expect different observers on any given $t = t_{\rm obs}$ hypersurface to construct different Hubble diagrams, even if they observe in the same direction, as they will not exist on the same hypersurface of homogeneity. By contrast, if we picked observers at different $t_{\rm obs}$, but the same value of the combination $t_{\rm obs} + Cr_{\rm obs}$, they would all be on the same homogeneous hypersurface, and would therefore construct identical Hubble diagrams.

We consider $100$ different observers at regular intervals of the $r$ coordinate throughout our averaging domain, on a hypersurface of constant  time, $t = t_{\rm obs} = 10\,$. In Fig. \ref{fig:farnsworth_raytracing} we show results for an observer located at $r_{\rm obs} = 2.5/\alpha$. As expected, there are no bumps in the functions $z(\lambda)$, $H^{\parallel}(z)$, $\Phi_{00}(z)$ and $\Psi_0(z)$, which in this case is entirely real. The effect of the Weyl curvature on the light ray's propagation, given by $\Psi_0$, is in all directions substantially smaller than the effect of Ricci curvature, $\Phi_{00}$. This is indicative of the late-time isotropisation of the Farnsworth metric, wherein it tends towards an FLRW universe with negative spatial curvature (and therefore no Weyl curvature).

\begin{figure}[t!]
\phantom{a}\hspace{-1.5cm}
    \begin{subfigure}{0.55\textwidth}
    \includegraphics[width=\linewidth]{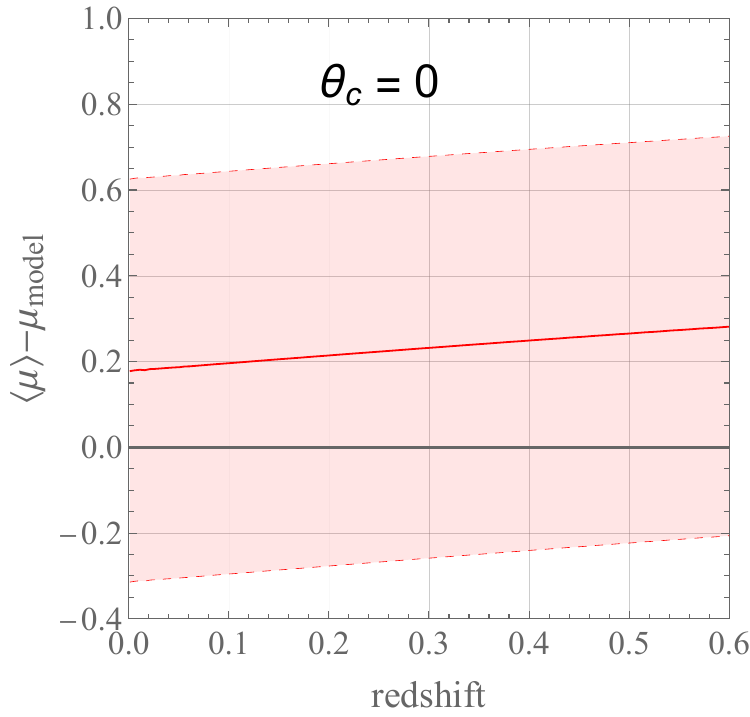} 
    \end{subfigure}
        \begin{subfigure}{0.55\textwidth}
        \includegraphics[width=\linewidth]{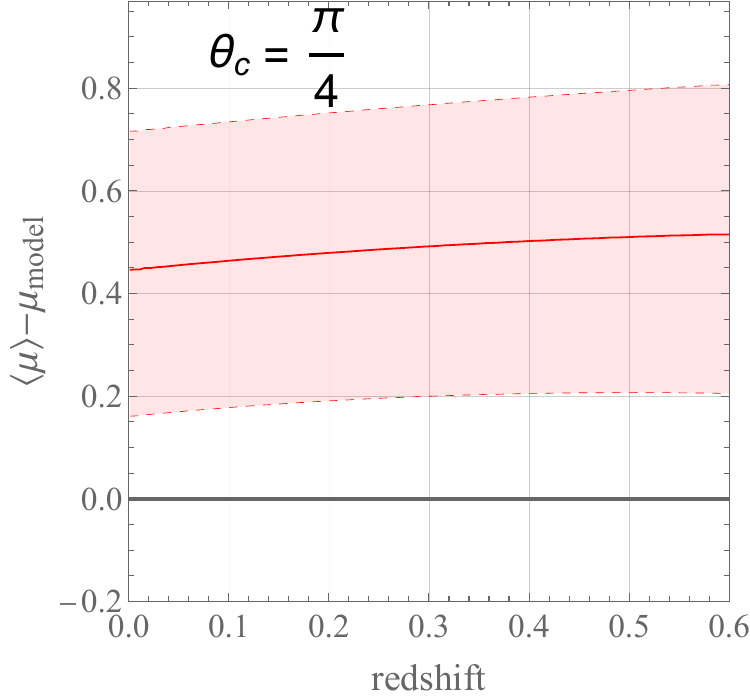}
    \end{subfigure}
    \caption{The ensemble mean of $\mu$ over 100 observers evenly spaced throughout the averaging domain in the Farnsworth geometry, compared to $\mu_{\rm model}$ obtained by mapping the scalar averages from that domain onto an LRS Bianchi type-$V$ model, for $\theta_c = 0$ (left) and $\pi/4$ (right).}
    \label{fig:Farnsworth_mean_mu_vs_model}
    \end{figure}

\subsection{Hubble diagrams}\label{subsec:farnsworth_hubble}

The Farnsworth cosmologies are of Bianchi type-$V$, with homogeneous surfaces tilted by a 3-velocity $Cf(r)/R'(t,r)$ with respect to the matter-comoving surfaces of constant time. It therefore seems natural to map the averages from the $t = {\rm cst.}$ surfaces to Bianchi type-$V$ cosmologies. 
The angular variation of $\mu$ at specified redshift shells, compared to this averaged Bianchi type-$V$ cosmology, is shown in Fig. \ref{fig:Farnsworth_mu-mu_model_theta}, where it can be seen that the difference between the curves increases with redshift at all angles.
This suggests that the model will fail to capture the correct Hubble diagram at intermediate and high redshifts, as verified in Fig. \ref{fig:Farnsworth_magnitude_vs_model}. This is in contrast to the case studied in Section \ref{sec:sinusoidal}, and we interpret it as being due to the lack of an homogeneity scale in the tilted hypersurfaces.

As expected, averaging over a large ensemble of observers does not alleviate the discrepancy, as displayed in Fig. \ref{fig:Farnsworth_mean_mu_vs_model}. Not only does $\avg{\mu} - \mu_{\rm model}$ not return to zero, but the standard deviation remains large in all cases. This indicates that not only does the lack of an homogeneity scale make the averaged Bianchi model perform poorly, but also that averaging in the constant-$t$ surfaces may not be a sensible procedure in the first place. This is because the tilt gives rise to a type of global inhomogeneity, meaning that imposing an homogeneous model gives rise to a Hubble diagram that bears no relation to one that observers in the space-time would record. 

\section{An $R' \neq 0$ Universe with Inhomogeneity}\label{sec:linear}

\begin{figure}[b!]
\phantom{a} \hspace{-1.5cm}
    \begin{subfigure}{0.53\textwidth}
    \includegraphics[width=\linewidth]{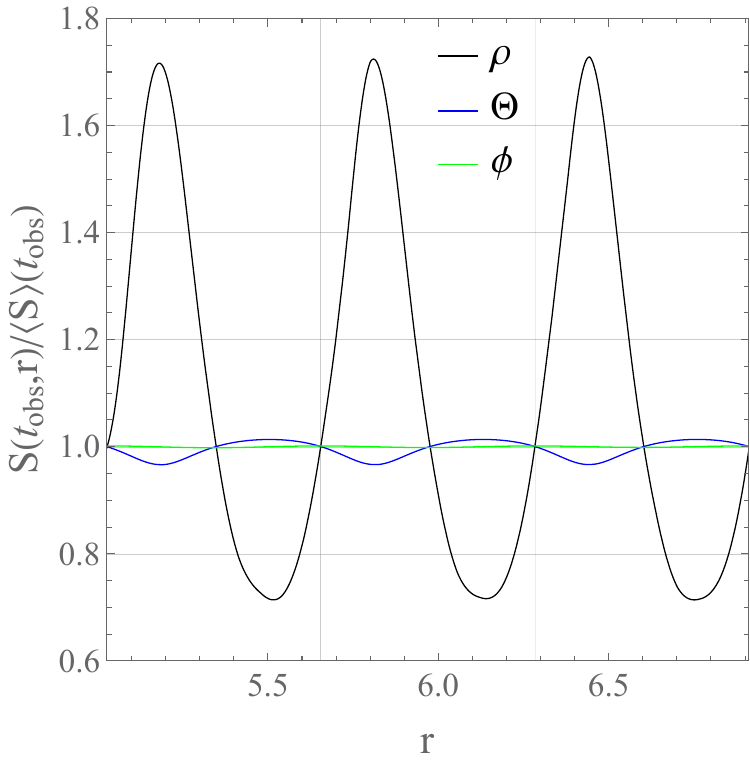}
    \caption{}
    \label{fig:density_linear}
    \end{subfigure}
    \begin{subfigure}{0.53\textwidth}
    \includegraphics[width=\linewidth]{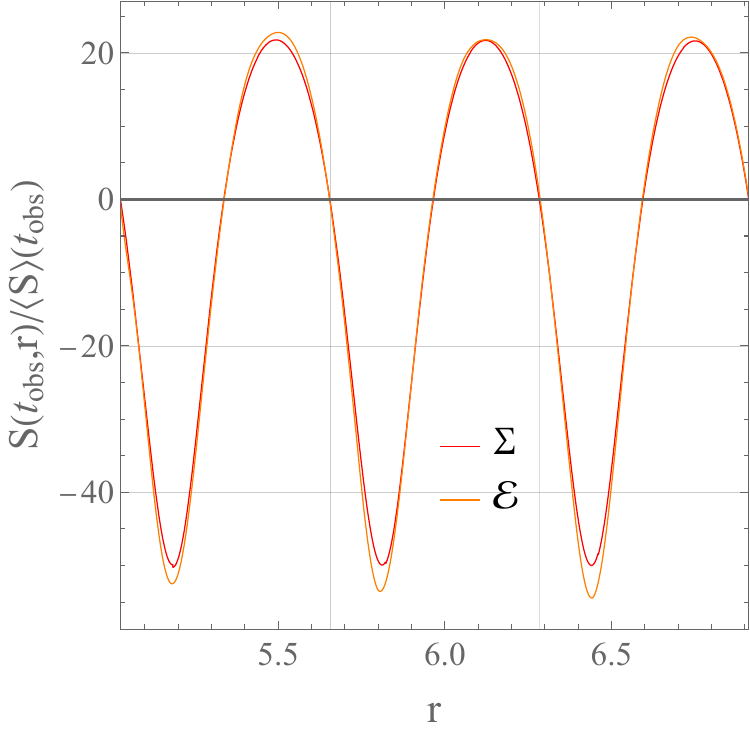} 
    \caption{}
    \label{fig:shear_linear}
    \end{subfigure}
    \caption{Left panel: matter density $\rho$, isotropic expansion $\Theta$, and space-like expansion $\phi$, at the observing time $t_{\rm obs}\,$, as a function of the $r$ coordinate in the $R' \neq 0$ models. Right panel: shear $\Sigma$ and electric Weyl curvature $\mathcal{E}$, also at $t_{\rm obs}$. All quantities are normalised by their averages, which have been calculated over the domain $\left[r_{\rm min}, r_{\rm max}\right)\,$. }
    \label{fig:linear_scalars}
\end{figure}

Finally, let us consider an inhomogeneous plane-symmetric universe with $R' \neq 0$. In this case the back-reaction scalars are not {\it a priori} restricted to be zero.  An homogeneous, tilt-free solution to the plane-symmetric metric with $R' \neq 0$ is provided by the metric functions in Eqs. (\ref{eq_R'_neq_0_metric_general}) and (\ref{eq_R'_neq_0_solution_parametric}) taking the form $f(r) = kr$, $m(r) = A f^3(r)$ and $t_0(r) = 0$, for some positive constants $k$ and $A$\,. To introduce inhomogeneity, we can therefore simply modify these functions, such that
\begin{eqnarray}\label{linear_model_definition}
    m(r) = A k^3 \,r^3\,,
    \qquad f(r) = k\left(r-b\sin^2{qr}\right)\,,\qquad {\rm and} \qquad 
    t_0(r) = 0\,,
\end{eqnarray}
where $b$ and $q$ are free parameters controlling the amplitude and frequency of oscillations in the inhomogeneities. 
The $1$+$1$+$2$-scalars, evaluated at the observing time $t_{\rm obs} = 40$, are displayed in Figs. \ref{fig:linear_scalars} for this case, where we have chosen $k = 5$, $A = 1$, and $q = 5$.

\begin{figure}[t!]
\phantom{a} \hspace{-0.cm}
\includegraphics[width=0.9\linewidth]{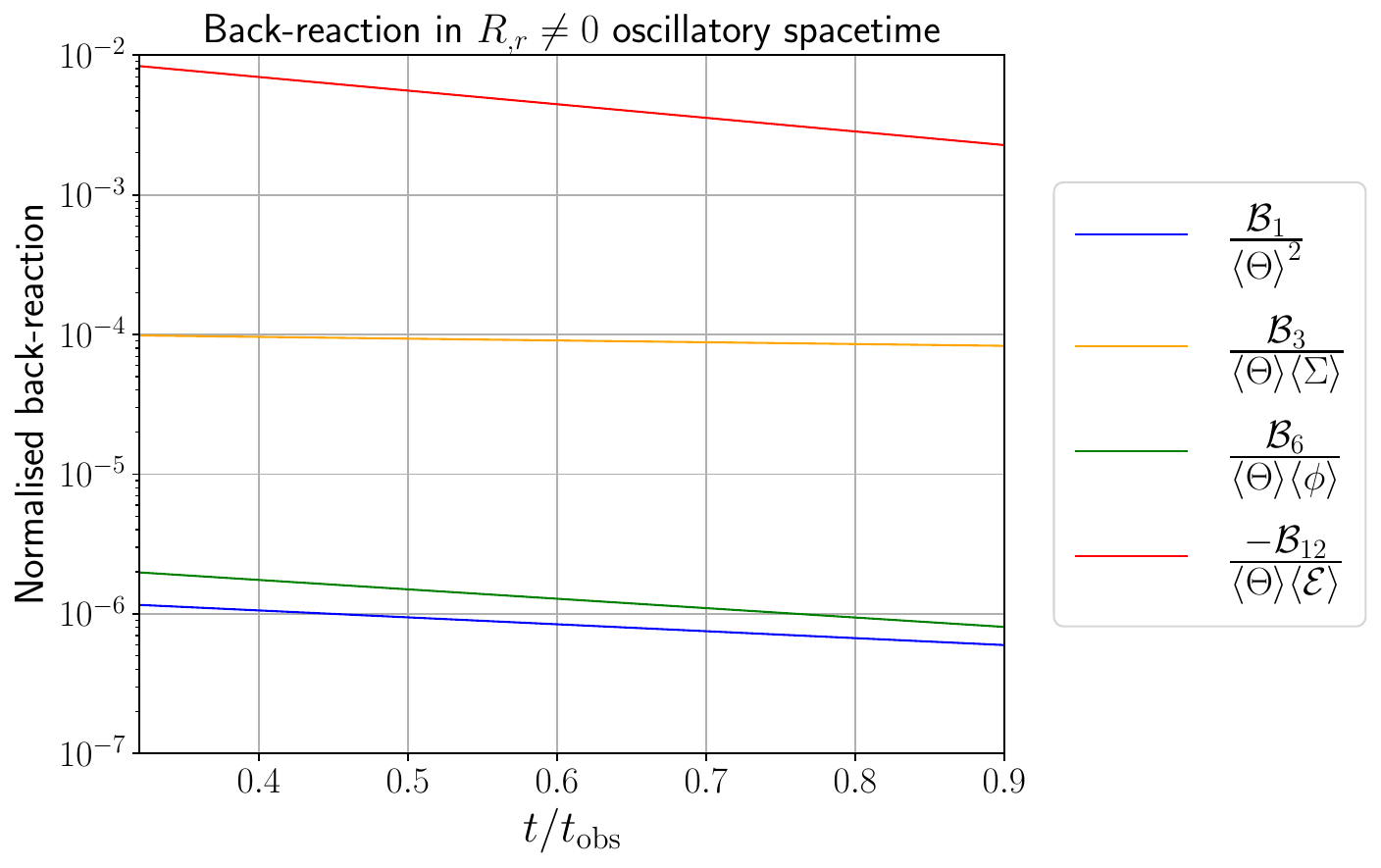}
\caption{Non-vanishing back-reaction scalars $\mathcal{B}_i$ for constant-time surfaces of the $R' \neq 0$ geometry. These scalars have all been made dimensionless, by normalising them with respect to the largest term from the equation in which they appear.
}
\label{fig:backreaction_linear}
\end{figure}

Importantly, we restrict the amplitude $b$ so that the matter density undergoes fluctuations of order unity, but is always non-negative. We find that $b = 0.1$ produces a density that is never less than $30\%$, or more than $180\%\,$, of its average value at $t=t_{\rm obs}$. 
This means that the model never reaches perfect matter-domination nor vacuum-domination (unlike in the $R' = 0$ case from Section \ref{sec:sinusoidal}), but that the density variations are still large. We then calculate the full set of scalar averages $\mathcal{B}_i$. 
By inspection of Fig. \ref{fig:linear_scalars}, one sees that $\Delta r = \pi/q$ defines the statistical homogeneity scale, meaning that one may choose any such oscillation cycle as constituting the extent of the $r$-coordinate in a well-motivated averaging domain (the $y$ and $z$ coordinates again being irrelevant, due to the plane symmetry of the space-time). We choose $r_{\rm min} = 9\pi/q$\,, so $r_{\rm max} = 10\pi/q\,$, and show our results in Fig. \ref{fig:backreaction_linear}. 

The reader can see that back-reaction scalars all make very small contributions to the evolution equations in this case. Their contributions are roughly constant in time, relative to the overall scale of the quantities in the equation, and only $\mathcal{B}_{12}$ (the back-reaction term in the evolution equation for $\avg{\mathcal{E}}\,$) is larger than $0.01\%$ of the dominant term in its evolution equation. Furthermore, $\mathcal{B}_{12}$ has the opposite sign to $\avg{\Theta}\avg{\mathcal{E}}$, so it is not causing the average Weyl curvature to grow, but rather is suppressing it. The other $\mathcal{B}_i$ are all of positive sign, but are negligibly small. Overall, one sees that although the back-reaction scalars are allowed to be non-zero in the $R' \neq 0$ plane-symmetric space-times, they are still highly restricted by symmetries. This is true even though the matter density fluctuations are of order unity, and seems to be independent of our precise choice of parameter values and functions.

\subsection{Ray tracing}

\begin{figure}[t!]

\phantom{a} \hspace{-1.5cm}
    \begin{subfigure}{0.53\textwidth}
    \includegraphics[width=\linewidth]{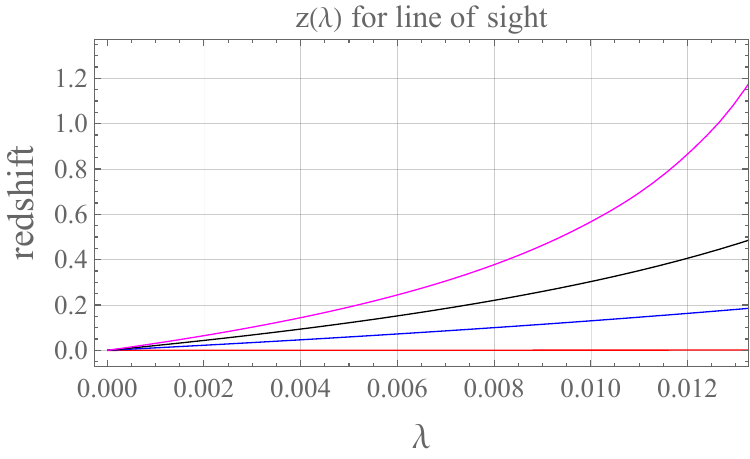}
    \caption{}
    \label{fig:z_lambda_linear}
    \end{subfigure}
    \begin{subfigure}{0.53\textwidth}
    \includegraphics[width=\linewidth]{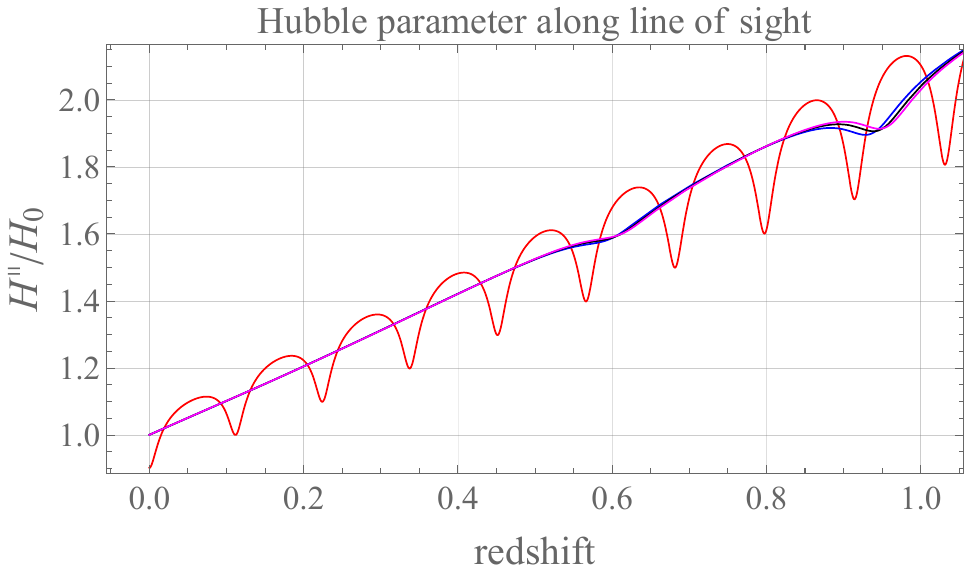} 
    \caption{}
    \label{fig:H_z_linear}
    \end{subfigure}
    \newline
    \phantom{a} \hspace{-1.9cm}
    \begin{subfigure}{0.55\textwidth}
    \includegraphics[width=\linewidth]{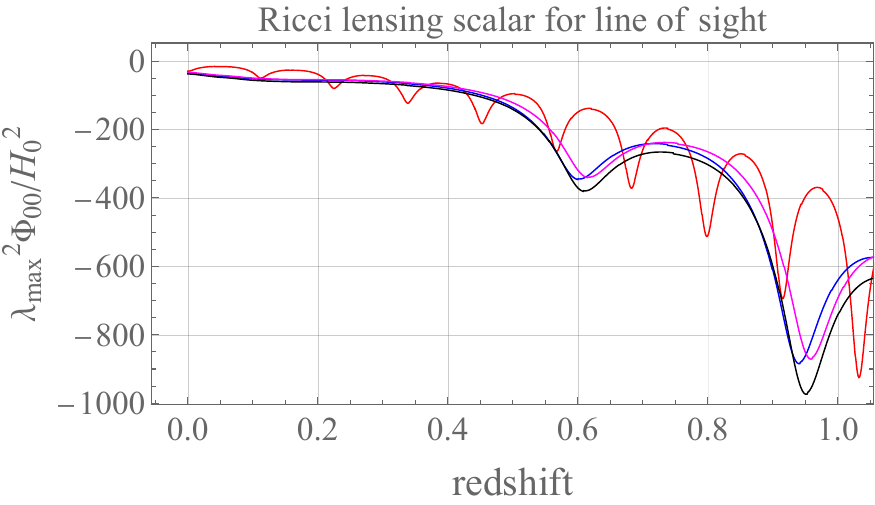} 
    \caption{}
    \label{fig:ric_lensing_linear}
    \end{subfigure}
    \begin{subfigure}{0.55\textwidth}
    \includegraphics[width=\linewidth]{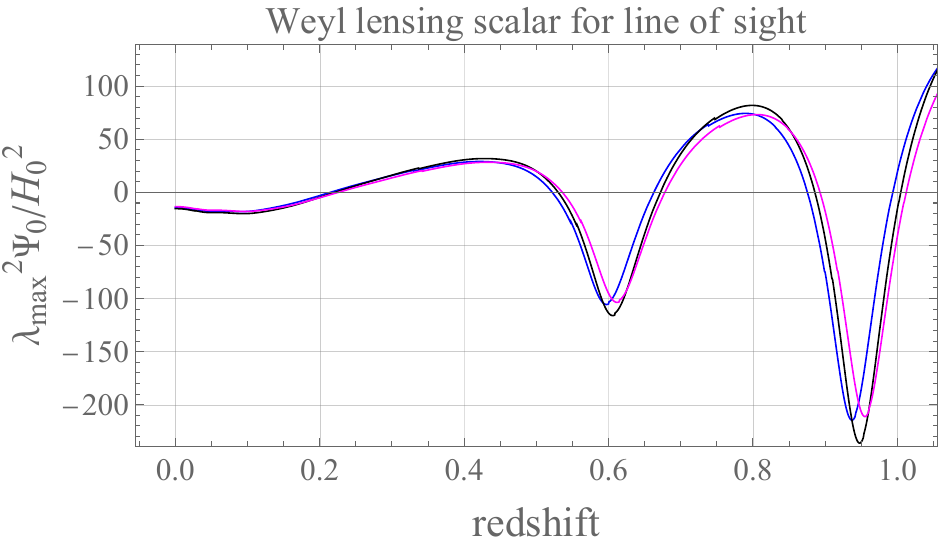}
    \caption{}
    \label{fig:weyl_lensing_linear}
    \end{subfigure}
    \caption{Top left: redshift as a function of affine distance in models with $R'\neq 0\,$, for observing angles $\theta_c = 0$ (red), $\pi/4$ (blue), $\pi/2$ (black) and $\pi/2$ (magenta) . Top right: line-of-sight Hubble parameter $H^{\parallel}$ as a function of redshift, normalised by its monopole at $z=0$. 
    Bottom left and right: Ricci and Weyl lensing scalars $\Phi_{00}$ and $\Psi_0$ as functions of redshift. Each of the Ricci and Weyl terms have been normalised by the maximum affine parameter value $\lambda_{\rm max}^{-2}$.}
    \label{fig:linear_raytracing}
\end{figure}

The results of our ray tracing procedure are summarised in Fig. \ref{fig:linear_raytracing}, where we have displayed $z(\lambda)$, $H^{\parallel}(z)$, $\Phi_{00}(z)$ and $\Psi_0(z)$ for null geodesics arriving at an observer at time $t=t_{\rm obs}$ and location $r_{\rm obs} = r_{\rm min} + \frac{\pi}{4q}$\,. 
Fig. \ref{fig:z_lambda_linear} shows that the redshift $z$ is monotonic in affine parameter $\lambda$ for all $\theta_c$ considered. 
The effect of inhomogeneities on the null rays is, however, clear once one considers the inferred line-of-sight Hubble parameter, which is displayed in Fig. \ref{fig:H_z_linear}. 
Here the oscillations in the metric function $f(r)$ are of amplitude $10\%$\,, so the effects of inhomogeneity on $H^{\parallel}$ are less drastic than in the $R'=0$ case (where the metric oscillated entirely between dust-dominated and vacuum regions).

\begin{figure}[b!]
\centering
    \includegraphics[width=0.6\linewidth]{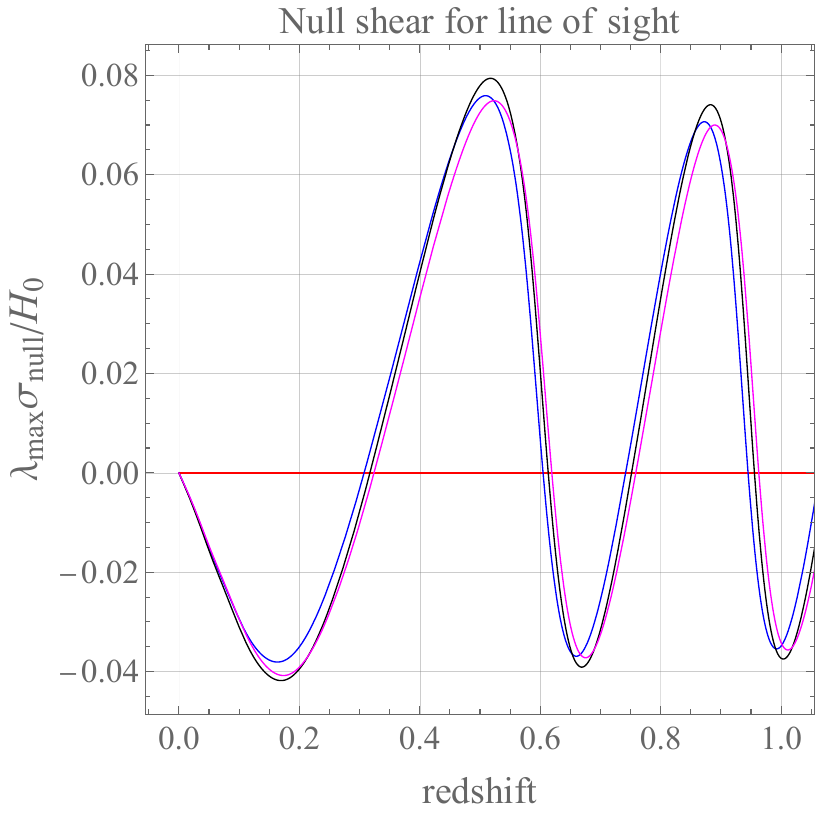}
\caption{Shear $\hat{\sigma}$ of the null congruence, as a function of redshift for models with $R'\neq0$, displayed for observing angles $\theta_c = 0$ (red), $\pi/8$ (blue), $\pi/4$ (black) and $\pi/2$ (magenta), and normalised by $\lambda_{\rm max}^{-1}$ in each case. We have considered an observer at $r_0 = r_{\rm min} + \pi/4q\,$, and at time $t=t_{\rm obs}$.}
\label{fig:linear_model_null_shear}
\end{figure}

The Ricci and Weyl curvature terms have the same oscillatory pattern as $H^{\parallel}$. 
Unless one is observing directly along the symmetry axis ($\theta_c = 0$), then the Weyl curvature contribution $\Psi_0$ is typically of the same order of magnitude as the Ricci contribution $\Phi_{00}$, which is displayed in Fig. \ref{fig:ric_lensing_linear}. Fig. \ref{fig:weyl_lensing_linear} shows that $\Psi_0$ has a changing sign in each case.
This means that for observations at high redshift, for which null rays will typically have to travel through many oscillations in the geometry, the effect of Weyl curvature will be suppressed relative to Ricci curvature, even though the two terms $\Psi_0$ and $\Phi_{00}$ are typically of comparable magnitude. 
This can be understood by studying Sachs' equation for the null shear (\ref{sachs2}), which integrates to
\begin{equation}\label{eq_solution_to_sachs2}
    \hat{\sigma}(z) = \frac{1}{d_A^2(z)}\int_0^z \frac{\mathrm{d}z' \, \Psi_0(z')\, d_A^2(z')}{\left(1+z'\right)^2 H^{\parallel}(z')}\,.
\end{equation}
As $H^{\parallel}$ is always positive in this case, the oscillations in the sign of $\Psi_0$ cause the right-hand side of Eq. (\ref{eq_solution_to_sachs2}) to remain bounded, when integrated to intermediate and high redshifts, as can be seen in Fig \ref{fig:linear_model_null_shear}. 
The effect of Weyl curvature on $d_A$ is communicated through the presence of $\bar{\hat{\sigma}}\hat{\sigma}$ in Sachs' equation for $d_A$. By $z = 1$, $\bar{\hat{\sigma}}\hat{\sigma}$ is four orders of magnitude smaller than $\Phi_{00}$, and hence the integrated effects of Weyl curvature are small.

\subsection{Hubble diagrams}\label{subsec:linear_hubble}

\begin{figure}[t!]
\phantom{a} \hspace{-1.75cm}
    \begin{subfigure}{0.53\textwidth}
    \includegraphics[width=\linewidth]{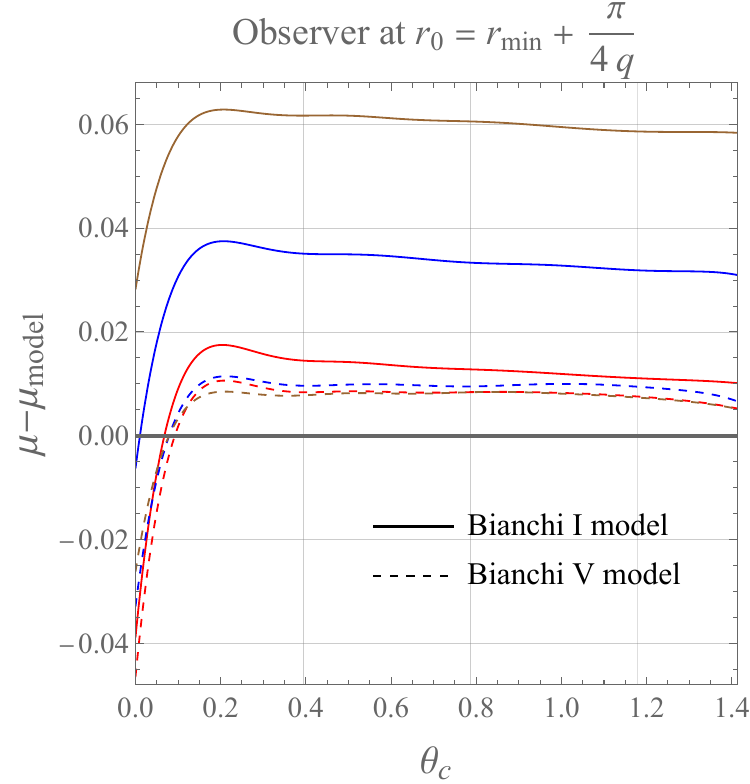} 
    \caption{}
    \label{fig:linear_model_mu-mu_model_theta}  
    \end{subfigure}
    \begin{subfigure}{0.53\textwidth}
    \includegraphics[width=\linewidth]{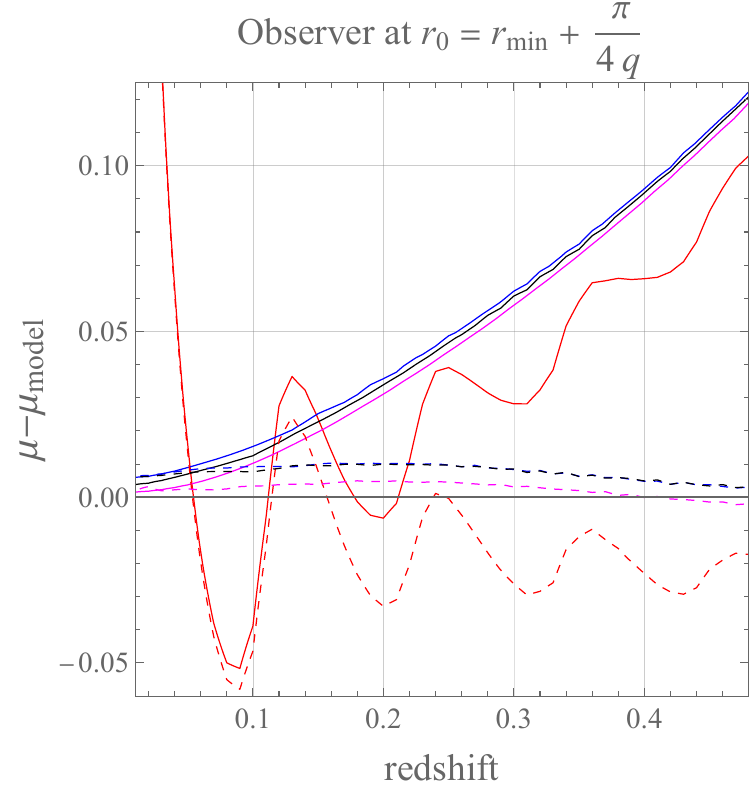} 
    \caption{}
    \label{fig:linear_magnitude_vs_model}
    \end{subfigure}
    \caption{Left: $\mu - \mu_{\rm model}$ as a function of $\theta_c$, for an observer at $r_{\rm min} + \frac{\pi}{4q}$ in the $R'\neq0$ models. Curves correspond to $z = 0.1$ (red), $z = 0.2$ (blue) and $z = 0.3$ (brown). 
    Right: $\mu - \mu_{\rm model}$ as a function of redshift, for the same observer. Curves are for $\theta_c = 0$ (red), $\pi/8$ (blue), $\pi/4$ (black) and $\pi/2$ (magenta). In both the left and right plots, $\mu_{\mathrm{model}}$ refers to averaged Bianchi type-$I$ (solid) and $V$ (dashed) cosmologies.}
\end{figure}

\begin{figure}[t!]
    \phantom{a} \hspace{-1.6cm}
    \begin{subfigure}{0.53\textwidth}
    \includegraphics[width=\linewidth]{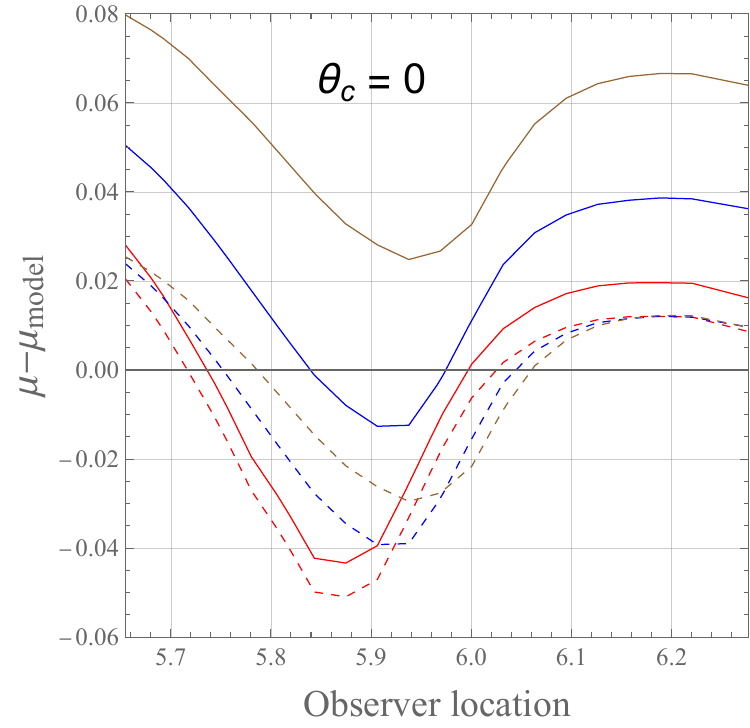} 
    \end{subfigure}
    \begin{subfigure}{0.53\textwidth}
    \includegraphics[width=\linewidth]{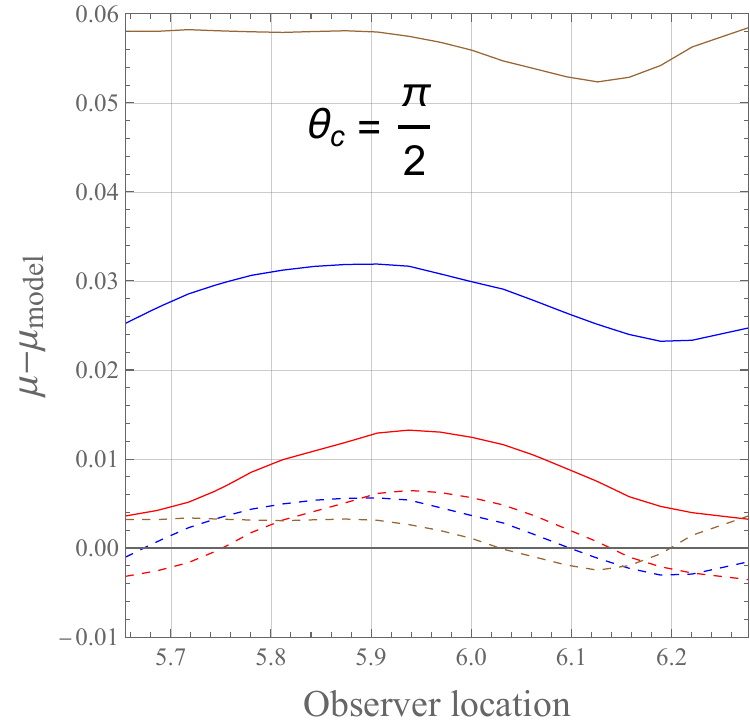} 
    \end{subfigure}
    \caption{The same set of plots as in Fig. \ref{fig:sinusoidal_mu-mu_model_vs_observer_location}, but for the $R'\neq 0$ models, and where $\mu_{\mathrm{model}}$ refers to averaged Bianchi type-$I$ (solid) and $V$ (dashed) cosmologies. Curves are for $z = 0.1$ (red), $z = 0.2$ (blue), and $z = 0.3$ (brown). The averaging domain extends from $r_{\rm min} = 9\pi/q$ to $r_{\rm max} = 10\pi/q\,$.}
    \label{fig:linear_mu-mu_model_vs_observer_location}
    \end{figure}
   
\begin{figure}[t!]
\phantom{a} \hspace{-1.5cm}
    \begin{subfigure}{0.53\textwidth}
    \includegraphics[width=\linewidth]{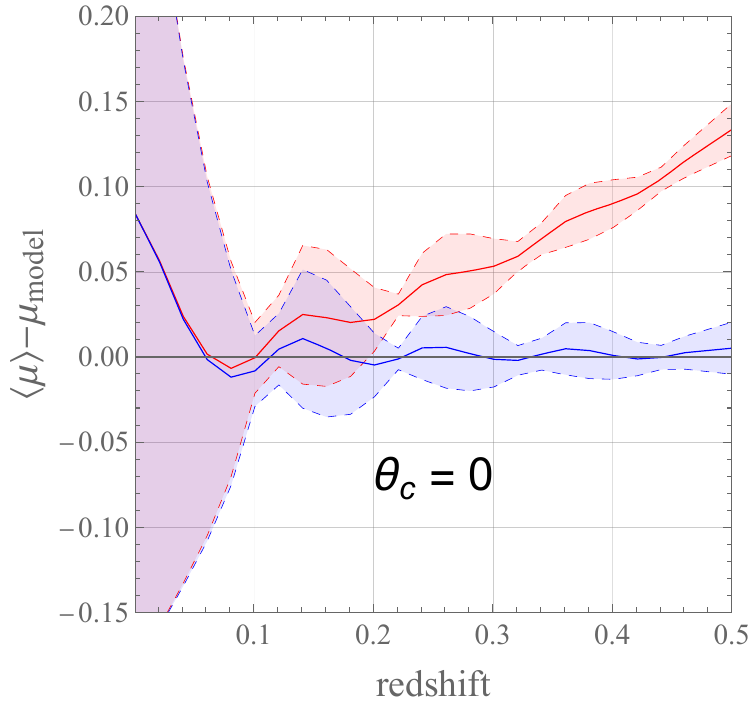} 
    \end{subfigure}
    \begin{subfigure}{0.53\textwidth}
        \includegraphics[width=\linewidth]{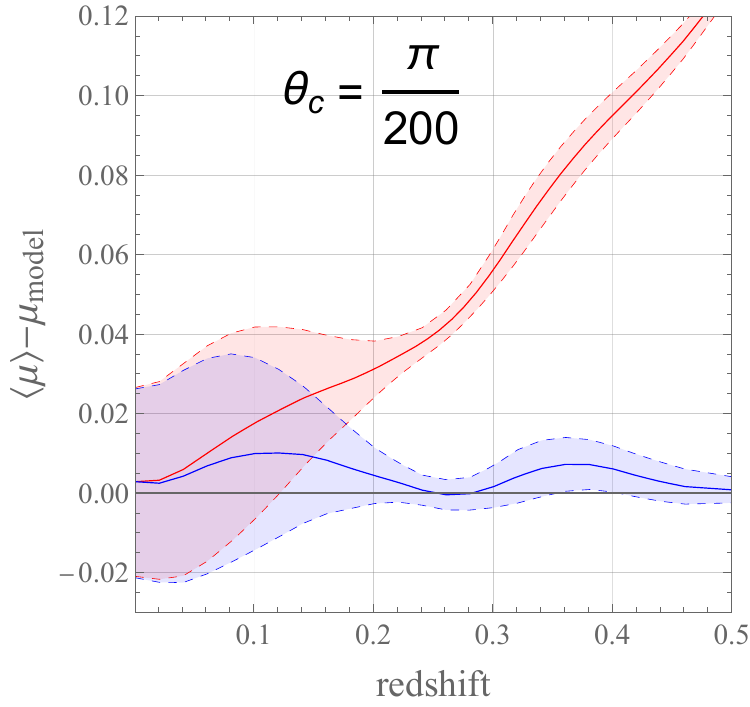} 
    \end{subfigure}
        \newline \phantom{a} \hspace{-1.5cm}
        \begin{subfigure}{0.53\textwidth}
        \includegraphics[width=\linewidth]{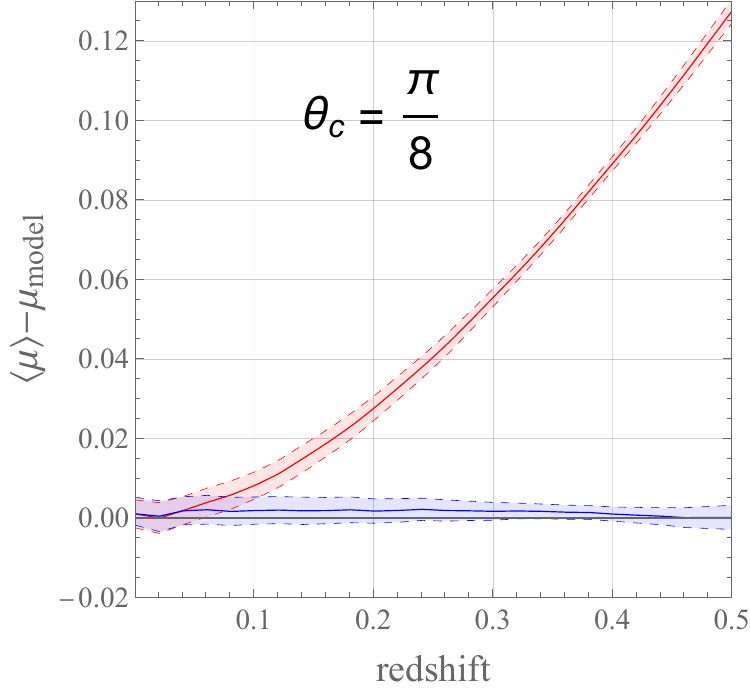} 
    \end{subfigure}
        \begin{subfigure}{0.53\textwidth}
        \includegraphics[width=\linewidth]{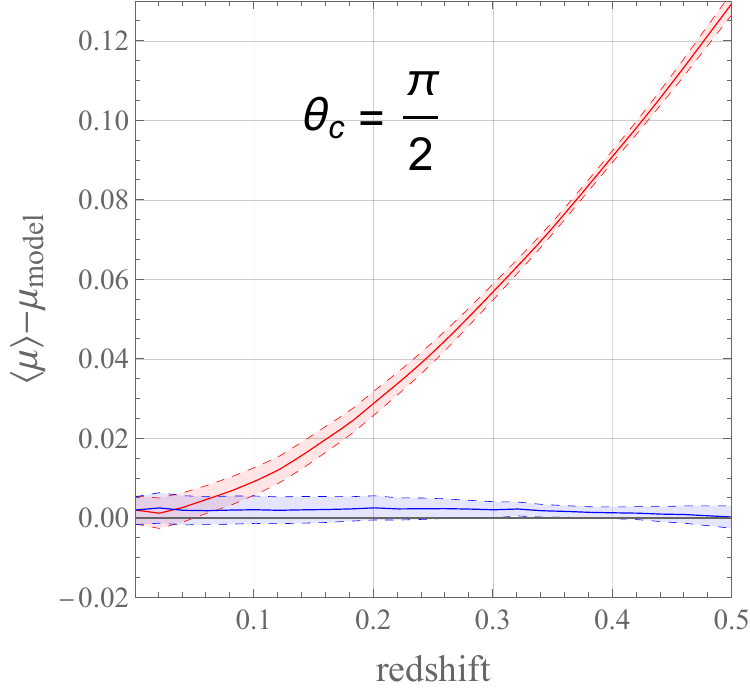} 
    \end{subfigure}
    \caption{$\avg{\mu} - \mu_{\rm model}$, calculated as a function of redshift in the $R'\neq0$ models. We have displayed results for $\theta_c = 0$, $\pi/200$, $\pi/8$ and $\pi/2$, and averaged over 100 observers. The $\mu_{\mathrm{model}}$ refers to averaged Bianchi type-$I$ (red), and $V$ (blue).}
    \label{fig:linear_mean_mu_vs_model}
    \end{figure}

In this section we consider both the Bianchi type-$I$ and $V$ geometries as candidate models, by mapping the scalar averages $\avg{\Theta}$, $\avg{\Sigma}$ and $\avg{^{(3)}R}$ onto these classes (as appropriate). Our averages are calculated over the domains described above, and evolved backwards in time from $t_{\rm obs}$ to $t_{\rm obs}/4\,$. As in the previous sections, we then perform ray tracing in those homogeneous averaged models by solving the geodesic equation for past-directed null geodesics emanating from observers at $t_{\rm obs}$, and compare the results of this to the results of ray tracing in the true inhomogeneous space-time. 

Fig. \ref{fig:linear_model_mu-mu_model_theta} shows a comparison of the distance moduli to the average models, as a function of $\theta_c$ for a given redshift. It demonstrates that the type-$V$ average model appears more appropriate than type-$I$, which is clear once one considers observing directions $\theta_c$ that are sufficiently far from the rotational symmetry axis. 
Not only are the differences $\mu - \mu_{\rm model}$ significantly smaller for type-$V$ than type-$I$, but they also decay with redshift.  However, at this point it is not yet clear that observations of type-Ia supernovae would actually indicate a preference for either average model, as any overall shift in $\mu$ can be removed by a recalibration of the intrinsic magnitude of the supernovae. All that would remain in that case would be the angular profile, which is not easily explained by either choice of homogeneous large-scale geometry. If this were all of the information available then one would not be able to select which average model would be more appropriate, which could lead to serious errors in inferring cosmological parameters.

Figure \ref{fig:linear_magnitude_vs_model} shows the Hubble diagrams $\mu(z)$ that would be constructed for observing directions $\theta_c = \left\lbrace 0, \pi/8, \pi/4, \pi/2\right\rbrace\,$, for redshifts up to $z\sim 0.5$. One sees that at low redshifts there is little distinction between the models, but at high redshift the Bianchi type-$V$ interpretation of the scalar averages is substantially preferred over type-$I$. 
In the case of the type-$V$ models, $\mu - \mu_{\rm model}$ remains very close to zero at all redshifts for $\theta_c = \pi/8$ and larger. This is a direct result of the angular profile of $\mu$ displayed in Fig. \ref{fig:linear_model_mu-mu_model_theta}: the function $\mu(\theta_c)$ is steepest near $\theta_c = 0$ and then rapidly becomes a nearly flat line just above the monopole. Therefore, $\mu - \mu_{\rm model}$ for each $\theta_c \gtrsim \pi/8$ is nearly flat.
This explains the lack of features in Fig. \ref{fig:linear_magnitude_vs_model}, for both observing locations and all the observing directions shown (except directly along the symmetry axis). On the other hand, the effect of inhomogeneities is clearest for $\theta_c = 0\,$, but there is still a relatively rapid trend towards zero, with $\vert \mu - \mu_{\rm model} \vert < 0.05$ for $z > 0.2$ for the type-$V$ models, in both plots. 
This shows the success of our averaging procedure at describing the large-scale Hubble diagram of an anisotropic cosmology, if the large-scale averaged model is chosen appropriately.

Let us now suppose that we had access to information from $100$ such observers, at different coordinate locations equally separated from one another, throughout our averaging domain (i.e. with $r_{\rm obs} \in \left[9\pi/q, 10\pi/q\right)$).  The effect of the inhomogeneity on distance measures in this case is displayed in Fig. \ref{fig:linear_mu-mu_model_vs_observer_location}, where it can be seen that the preference for the type-{\it V} model only becomes clear at redshifts $z \gtrsim 0.2\,$. 

Performing an ensemble average over this set of observers gives the results displayed in Fig. \ref{fig:linear_mean_mu_vs_model}, which shows that there are small effects in all directions in the redshift range $z < 0.1$. This is due to some observers receiving photons that have just moved through a region of high density, and so a large negative Ricci curvature term $\Phi_{00}$, and others receiving photons coming through underdensities. These local effects are, however, only pronounced for $\theta_c$ close to zero, as can be seen from the results for $\theta_c = {\pi}/{200}$, which show a rapid transition in behaviour as the observing angle $\theta_c$ is increased from zero. We also note that oscillatory features in the Hubble diagram are highly suppressed if one is observing away from the symmetry axis. This is in obvious contrast to the $R'=0$ models considered in Section \ref{sec:sinusoidal}, where Fig. \ref{fig:sinusoidal_mean_mu_vs_model} shows that for the $R' = 0$ model, the ``average Hubble diagram'' has distinct features of inhomogeneity for $\theta_c = \pi/8$ and $\pi/4$ as well as just $\theta_c = 0$. 

While the $\theta_c = 0$ and $\theta_c = {\pi}/{200}$ cases demonstrate significant effects from the inhomogeneities, it remains true that $\avg{\mu}$ is always consistent with $\mu_{\rm model}$ to within $1\sigma$. It can also be seen that the size of the oscillations, and their associated confidence intervals, is clearly decaying with increased redshift. Finally let us note that the offset at $z = 0$ can be explained with reference to the Hubble parameter, since for $z \ll 1$ a Taylor series expansion of $d_L(z)$ shows that the leading term at low redshifts (in a generic space-time) is given by $d_L \simeq z/H_0^{\parallel}$ \cite{Heinesen_2021}, which in our diagrams leads to a vertical displacement in $\mu \sim \log{d_L}$. At the same time, the ensemble average of the line-of-sight Hubble parameters $H^{\parallel}_0$ that are measured by each of the observers (as displayed in Fig. \ref{fig:H_z_linear}) is not always equal to the line-of-sight Hubble parameter in the averaged homogeneous model, which means that a local effect on $\mu$ is entirely expected. In the $\theta_c = \pi/8$ and $\pi/2$ diagrams we have subtracted off the small offset. Thus, the nearly flat line $\avg{\mu} - \mu_{\rm model}$ that appears for $z \gtrsim 0.1$ is consistent with zero, which reflects the fact that a flat vertical displacement would not be observable.  
This concludes our study of the Hubble diagrams constructed by observers in statistically homogeneous, anisotropic cosmologies, and their ability to be fit by the models that result from our averaging formalism.

\section{Discussion}\label{sec:discussion}

We have constructed Hubble diagrams in universes that are inhomogeneous, and anisotropic on large scales. These diagrams depend on the line-of-sight of the particular observer, as well as their position in space-time. We have then compared these diagrams to those that would be created by considering the large-scale average space-time, using a formalism we presented in Ref. \cite{Anton_2023}.
In order to carry out this comparison, we have focused on three families of cosmological models within the plane-symmetric class of dust-dominated solutions of Einstein's equations. These solutions admit closed-form exact solutions, and for arbitrary amounts of inhomogeneity to be introduced in the directions orthogonal to the surfaces of symmetry (though care is needed to avoid situations that may involve shell-crossings and singularities). The homogeneous sub-class of this set of solutions belong to the LRS Bianchi type-$I$ and $V$ cosmologies, which are therefore considered to be the ``target space'' for the averaged cosmological models.

While the observations made by any one observer in these space-times are not necessarily reproduced well by the averaged cosmological models, we find that the observations made by many observers have an average that can be described well. In particular, we found in Sections \ref{sec:sinusoidal} and \ref{sec:linear} that there exist averaged anisotropic cosmological models that can accurately predict the Hubble diagram for average observations made in all directions to within $1\sigma$.
This is a non-trivial result, as the Hubble diagrams can take very different shapes in different directions in universes that exhibit large-scale anisotropy. In particular, it extends previous studies that studied only observations along lines-of-sight that are aligned with principal null directions \cite{Bull_2012}.
Furthermore, the $1\sigma$ confidence interval generically shrinks as redshift increases, meaning that the average model becomes a better approximation for typical observers who make measurements over larger distances, as one would hope for a useful cosmological model.

While our study has shown some successes for the anisotropic cosmologies that result from the application of our averaging formalism, it has also shown some clear warning signs. In particular, it is clear that the choice of foliation on which the averaging is performed must be chosen with care. This is exemplified by the tilted Farnsworth cosmologies studied in Section \ref{sec:farnsworth}, where we found that an anisotropic homogeneous model constructed from our averaging procedure could {\it not} fit the Hubble diagrams of observers accurately, even in an average sense.
This is despite the fact that the Farnsworth space-time is genuinely spatially homogeneous, and clearly indicates the importance of suitably foliating the space-time. In general, we do not expect averaged cosmological models to reproduce the average of observables if there is no statistical homogeneity scale. If a foliation is such that no scale of this type exists, then we expect that choice to fail.

Another area which requires care, in order to get a sensible result for the averaged cosmology, is the choice of target space for the symmetries of the averaged model. In particular, if the averaged model does not allow for all aspects of the averaged covariant scalars to be accounted for, then it is unlikely to reproduce the average of observations made within the space-time. We believe this to be the reason why the Bianchi type-$I$ models failed to reproduce the average Hubble diagrams for the observers considered in Section \ref{sec:linear}. In that case the average spatial curvature is non-zero, and so it needs an averaged cosmology that allows for this possibility to exist. This is the case for Bianchi type-$V$ models (which reproduced observables well), but not for Bianchi type-$I$ (which did not, at high $z$). We suspect the same will be true for cosmological models with large-scale tilt, which is only allowed in a restricted set of Bianchi classes.

Ultimately, as cosmologists, we would like to obtain a full picture of the large-scale properties of our Universe. As the number of type-Ia supernovae, quasars, radio galaxies and other distant sources we observe rises in the coming years, it will be possible to construct an increasingly precise sky map of $d_L(z)$, to $z \sim 1$ and beyond. If these observations continue to support the existence of large-scale anisotropies in the Universe, then we will need cosmological models that can include that freedom. 
The analysis presented in this work constitutes a step towards understanding the Hubble diagram in such space-times, which we hope will be of use for understanding the theoretical modelling problem at hand, as well as potentially providing a framework within which anisotropic observations in the real Universe can be understood and interpreted. In future work we hope to investigate situations in which our back-reaction scalars can be large, while maintaining statistical homogeneity within our averaging domains. 
This did not prove possible in the plane-symmetric space-times we considered in the present work, but could potentially be found in more general situations.
The use of relativistic simulations \cite{macpherson2021luminosity, Macpherson:2022eve, Adamek_2018, lepori2020weak, lepori2021cosmological} may well prove to be fruitful in this regard.

\appendix

\flushleft
{\bf Acknowledgements:} We are grateful to Charles Dalang, Pierre Fleury, Asta Heinesen and Phil Bull for helpful discussions. The authors acknowledge support from the Science and Technology Facilities Council (STFC) grant ST/P000592/1.

\bibliographystyle{unsrt}

\end{document}